# Action needed to make carbon offsets from tropical forest conservation work for climate change mitigation

Thales A. P. West[1,2,*], Sven Wunder[3,4], Erin O. Sills[5], Jan Börner[6,7], Sami W. Rifai[8], Alexandra N. Neidermeier[1], Andreas Kontoleon[2,9]

[1] *Environmental Geography Group, Institute for Environmental Studies (IVM), VU University Amsterdam, Amsterdam, The Netherlands*
[2] *Centre for Environment, Energy and Natural Resource Governance, University of Cambridge, Cambridge, United Kingdom*
[3] *European Forest Institute (EFI), Barcelona, Spain*
[4] *Center for International Forestry Research (CIFOR), Lima, Peru*
[5] *Department of Forestry and Environmental Resources, North Carolina State University, Raleigh, USA*
[6] *Center for Development Research (ZEF), University of Bonn, Bonn, Germany*
[7] *Institute for Food and Resource Economics (ILR), University of Bonn, Bonn, Germany*
[8] *ARC Centre of Excellence for Climate Extremes, University of New South Wales, Sydney, Australia*
[9] *Department of Land Economy, University of Cambridge, Cambridge, UK*

*Corresponding author: t.a.pupowest@vu.nl

## Summary

Carbon offsets from voluntarily avoided deforestation projects are generated based on performance vis-à-vis *ex-ante* deforestation baselines. We examined the impacts of 27 forest conservation projects in six countries on three continents using synthetic control methods for causal inference. We compare the project baselines with *ex-post* counterfactuals based on observed deforestation in control sites. Our findings show that most projects have not reduced deforestation. For projects that did, reductions were substantially lower than claimed. Methodologies for constructing deforestation baselines for carbon-offset interventions thus need urgent revisions in order to correctly attribute reduced deforestation to the conservation interventions, thus maintaining both incentives for forest conservation and the integrity of global carbon accounting.

## Introduction

For nearly two decades, the performance-based payment mechanism for reduced carbon emissions from deforestation and forest degradation known as REDD+ has been under intense debate (Angelsen, 2017). While regulations and capacity for national REDD+ programs are still under development (Börner et al., 2018; FAO, 2019), many stand-alone, voluntary REDD+ projects are operational worldwide (Atmadja et al., 2022). These projects intend to conserve forests through multiple activities, like improved monitoring and control, promotion of sustainable practices, and local stakeholder engagement, often funded by the commercialization of carbon offsets (each corresponding to 1 Mg $CO_2$ either removed from or not emitted to the atmosphere). In 2021 alone, two-thirds of the 227.7 million offsets from the land-use sector (excluding agriculture) traded in environmental markets, with a total value of USD 1.3 billion, originated from REDD+ projects (Donofrio et al., 2022).

Numerous policy discussions and initiatives currently focus on how to scale and integrate the carbon emission reductions claimed by voluntary carbon-offset projects, particularly from REDD+ activities, into climate policies and Nationally Determined Contributions (NDCs) reported to the UNFCCC (FAO, 2019; Lee et al., 2018; Taskforce on Scaling Voluntary Carbon Markets, 2021; Verra, 2021a; Voluntary Carbon Markets Integrity Initiative, 2021). However, there is little rigorous evidence on the contributions of this type of initiative (Duchelle et al., 2018; Sills et al., 2017), with some studies suggesting that many are associated with little or no actual emission reductions (Badgley et al., 2021; Calel et al., 2021; Cames et al., 2016; Haya et al., 2020; Kollmuss et al., 2015; West et al., 2020).

Carbon offsets from REDD+ projects are issued based on the comparison between the observed forest cover in the project sites and deforestation baseline scenarios expected to have been realized in the absence of REDD+, which remain *de facto* unobservable (FAO, 2019). Many project baselines are informed by extrapolation of historical deforestation averages or trends (West et al. 2020). These crediting baselines may become unrealistic counterfactuals with changes in economic or political conditions influencing deforestation (Busch and Ferretti-Gallon, 2017; West and Fearnside, 2021). But baselines could also be opportunistically inflated by profiteers seeking to financially benefit from selling as many offsets as possible, even when these lack environmental additionality, i.e., are unlikely to reflect actual emission reductions (Rifai et al., 2015; Seyller et al., 2016; Wunder, 2007).

This study provides a pan-tropical comparison between *ex-post* deforestation counterfactuals, informed by observable control areas, and the *ex-ante* baselines adopted by 27 voluntary REDD+ projects in six tropical countries: Peru, Colombia, Democratic Republic of Congo (DRC), Tanzania, Zambia, and Cambodia (Tables S1 & S2; Fig. 1 & S1) certified under the Verified Carbon Standard (Verra, 2019). Because some projects are composed of multiple disconnected sites, we evaluate those areas individually, increasing our sample to 31 sites. We present both project-specific and cross-project analyses based, respectively, on the standard and generalized versions of the synthetic control (SC) method for causal inference, to estimate reductions in deforestation in project sites attributable to the REDD+ interventions (Abadie et al., 2021; Xu, 2017). The SCs were constructed based on selected control areas ("donors") exposed to similar levels of deforestation pressure (as measured by the average annual deforestation in the projects' 10-km buffer zones prior to the project implementation) and with similar characteristics as the REDD+ sites, from project-specific donor pools, which combined, could replicate the historical deforestation pattern in the project areas (see SI for details). Before interpreting results, we conducted project-specific "validation" tests to check whether the standard SC method was able to construct SCs with similar deforestation rates as project areas during the immediate pre-project period (West et al., 2020). Conservatively, we focus the discussion of our results on the projects with SCs that performed well in the validation test, i.e., with gaps between the project and its SC deforestation at the end of the validation period lower than 0.5% of the size of the project area (Tables S3 & Fig. S2). To address the concern that the selected control areas may not have the same chance to be selected as REDD+ project sites, we also estimated REDD+ impacts on deforestation based on the comparison of operational project sites with "yet-to-become" project areas throughout the study period using the matching-based methods for time-series cross-sectional data developed by Imai et al. (2018). Because the evaluated projects span multiple countries and contexts, our analyses can shed light on the robustness of the assumptions adopted for the construction of REDD+ baselines under a wide range of deforestation conditions (Fig. S3).

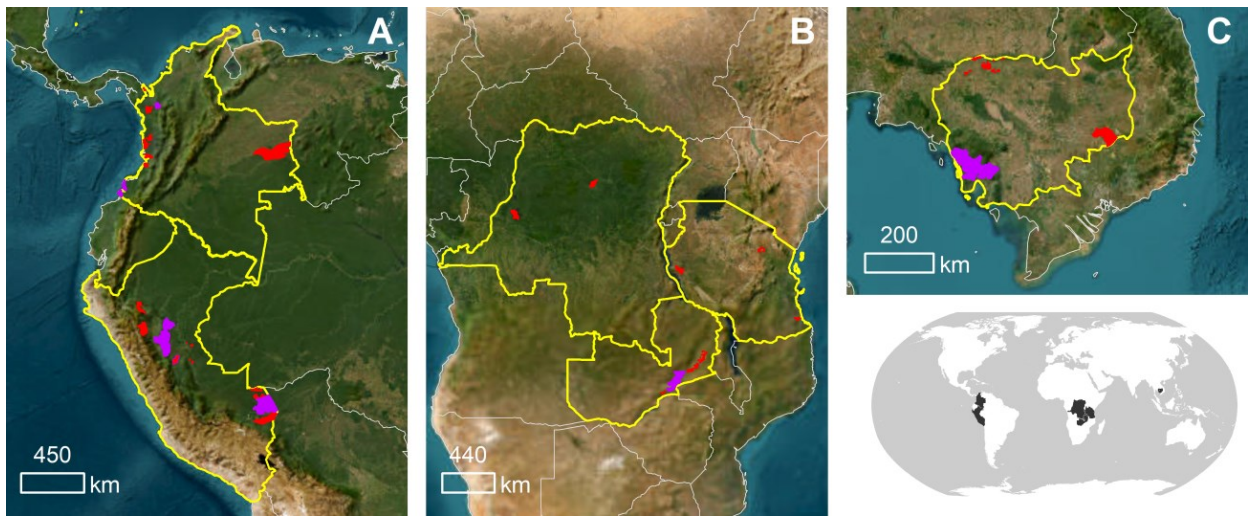

Figure 1. Voluntary REDD+ project sites included in the study (red areas) from Peru and Colombia (A), Democratic Republic of Congo, Tanzania, and Zambia (B), and Cambodia (C). Purple areas are the sites excluded from the analysis.

### Individual REDD+ project impacts

The individual SC analyses show mixed impacts of the voluntary REDD+ projects on deforestation. Results from the validation tests suggest that the SC method could replicate relatively well pre-REDD+ deforestation trends in 29 of the 31 "to-be" project sites (Tables S1 & Fig. S2). After constructing the SCs for the 31 sites, we discarded one project (#1775-1) from the analyses due to the poor fit between the deforestation in the SC and the REDD+ site prior to project implementation. Four other projects (#985, #1360-1, #1389, and #1748) were also discarded because of substantial disagreements between the REDD+ sites' and the SCs' buffer deforestation during the pre-project period. Our final sample was thus reduced to 26 project sites.

Six of the evaluated 26 project sites showed some evidence of additional reductions in deforestation compared to their individual SCs, although generally not to the extent claimed by the projects based on their crediting baselines (Fig. 2, S4, & S5). Additionality was most likely in Peru, where half of the REDD+ sites had significantly less deforestation than the *ex-post* counterfactuals, with statistical significance confirmed by placebo tests. Only one of the seven Colombian project sites, and one of two Cambodian sites, achieved significant deforestation reductions according to the SCs and placebo tests. No evidence of avoided deforestation was found for the REDD+ sites in the DRC, Tanzania, and Zambia vis-á-vis their counterfactuals.

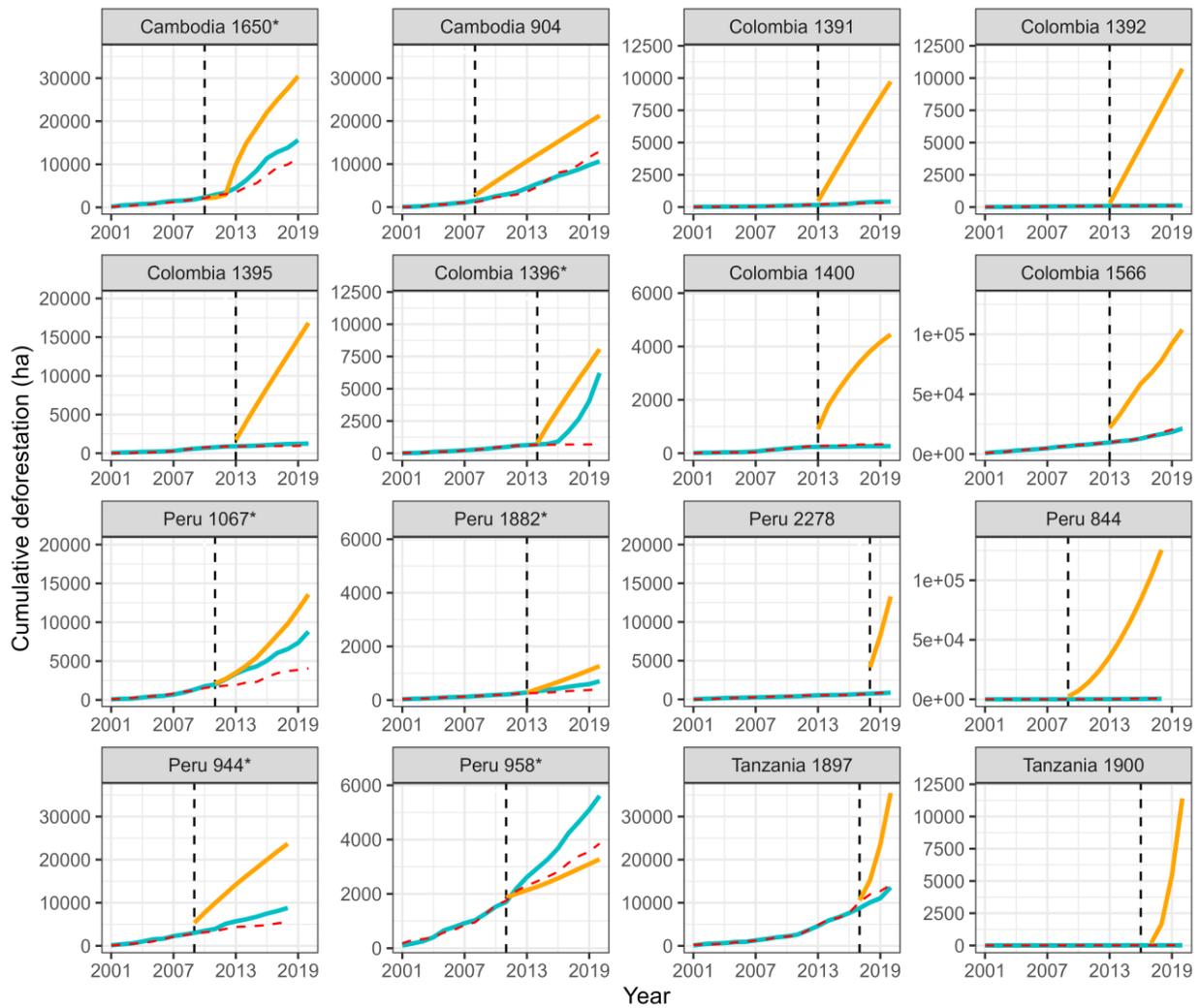

Figure 2. Cumulative post-2000 deforestation in the REDD+ project sites (dashed red line) and synthetic control (SC) areas (solid blue line) versus the baseline scenarios adopted by the REDD+ projects (solid orange line). Dashed black lines indicate the project implementation year. Asterisks indicate a significant reduction in deforestation in the REDD+ site compared to the SC (note: scales differ).

**Average REDD+ project impacts**

Average project impacts on deforestation (i.e., average treatment effects on the treated; ATT) in Peru, Colombia, and Africa (DRC, Tanzania, and Zambia) were estimated with the generalized SC (GSC) method (Fig. 3, upper panels). Cambodian projects were excluded from this analysis, due to the limited sample size. Unlike the individual project evaluations, the GSC analyses were based exclusively on annual deforestation rates and time-variant covariates. Such a distinction between the two methods increases the robustness of our results. The GSC analyses were based on two independent sets of selected control areas for each region: in the first set, only the donors selected for the construction of the individual SCs were considered, whereas in the second set, controls were selected based on a genetic matching technique (Diamond and Sekhin, 2013), independent of the SC analyses.

For the former set of selected controls, the average impact of the Peruvian projects on forest loss was −0.22% or 686 ha year$^{-1}$ (p-value = 0.10; Table S4). A similar ATT size was found for the African projects (−0.20% or 412 ha year$^{-1}$; Table S5), whereas a smaller effect was associated with the Colombian projects (−0.02% or 49 ha year$^{-1}$; Table S6). However, the estimates for both the Colombian and African projects were not significant (p-values of 0.61 and 0.26, respectively). Even assuming the estimated average reductions in deforestation to be significant in all three countries (a plausible assumption given our small sample sizes; Tables S4–S6), they would still be substantially lower than the average baseline deforestation rates adopted by the projects from Peru (3661 ha year$^{-1}$), Colombia (2550 ha year$^{-1}$), and Africa (2700 ha year$^{-1}$) through 2020. The Peruvian projects on average reduced deforestation in the REDD+ sites within the first four years following REDD+ implementation in comparison to the GSC (Fig. 3, lower panels). The GCSs indicate no significant reductions from projects in Colombia and the combined African countries.

These results are robust to using control areas selected with the genetic matching technique. Based on those controls, we estimate ATTs of the Peruvian, Colombian, and African REDD+ sites as −0.42% (1269 ha) year$^{-1}$, −0.05% (137 ha) year$^{-1}$, and −0.13% (258 ha) year$^{-1}$, respectively (Tables S7–S9 & Fig. S6, upper panel); again, only the Peruvian estimate was significant (p-values of 0.01, 0.34, and 0.33, respectively). Also similarly, the GSCs suggests that some average reduction in deforestation was achieved in Peru, but again restricted to the first four years of the project (Fig. S6, lower panels), with no significant reductions observed in Colombia or in the African countries.

Finally, results from the comparison between already operational and "not-yet-operational" project sites corroborate the findings from the GSC analyses. The average REDD+ impacts on deforestation ranged from −0.06% year$^{-1}$ to 0.10% year−1 (or −92 ha year$^{-1}$ to 103 ha year$^{-1}$), across all model settings, but none of the estimates were statistically significant (Fig. S7).

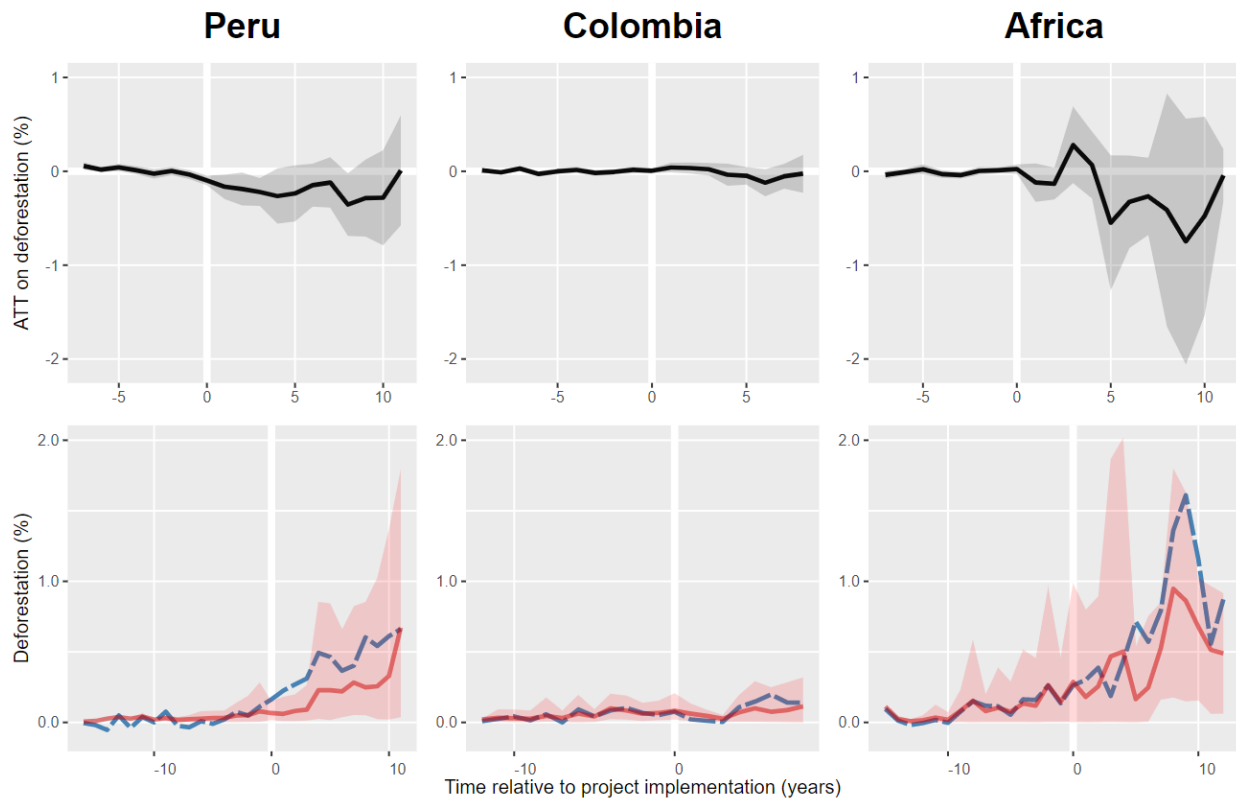

Figure 3. Estimated average impacts of REDD+ projects from Peru, Colombia, and Africa on annual deforestation, using the generalized synthetic control (GSC) method and selected controls from the individual synthetic control analyses. Upper panels display the average treatment effect on the treated (ATT) project sites. Lower panels display projects' (solid red line) and counterfactuals' (dashed blue line) deforestation averages. Shaded red areas represent bootstrapped 95% confidence intervals around the projects' deforestation average.

## Carbon offset implications

We investigated the implications of our findings for the environmental integrity of the credits issued by the REDD+ projects. These implications are based on the 18 out of 27 sampled projects with sufficient publicly available information about baseline deforestation rates (Table S10; Fig. 2 & S8). According to the projects' *ex-ante* estimates, up to 89 million carbon offsets could potentially have been generated by the REDD+ projects from our sample until 2020. Yet, 63.2 million of these offsets (71%) would have originated from projects that have not significantly reduced deforestation (and emissions) compared to their SCs. The remaining 25.8 million offsets (29%) would have originated from projects likely associated with some avoided deforestation, but not to the extent expected by the project developers. If we replace the *ex-ante* baselines adopted by the projects with the deforestation observed from the SCs, our estimates suggest that only 5.5 million (6.2%) of the 89 million *ex-ante* offsets from the REDD+ projects would likely be associated with additional carbon emission reductions.

As of November 2021, those 18 REDD+ projects have issued 62 million carbon-offset credits (Table S2). Out of those, at least 14.6 million (24%) have already been used by individuals or organizations around the world to offset their greenhouse gas emissions. Thus, according to our SC-based estimates, these projects have already been used to offset almost three times more carbon emissions than their actual contributions to climate change mitigation—with another 47.4 million carbon offsets being readily available in the market.

**Discussion**

Overall, the weight of evidence suggests that voluntary REDD+ projects in our sample across six tropical countries achieved much less avoided deforestation than anticipated by project developers. Only a few projects achieved significant reductions in comparison to the *ex-post* counterfactuals. Our findings corroborate prior studies questioning the additionality, and thus environmental integrity, of this type of carbon-offset intervention (Badgley et al., 2021; Calel et al., 2021; Cames et al., 2016; Haya et al., 2020; Kollmuss et al., 2015; Seyller et al., 2016). Exaggerated baseline scenarios are the biggest part of the underlying problem; unexpressive conservation performance by the REDD+ projects supplements the picture.

In an evaluation of REDD+ projects in the Brazilian Amazon, West et al. (2020) pointed to the potential confounding effect created by Brazil's post-2004 policy interventions to control deforestation, triggering a substantial reduction in forest loss between 2004 and 2012 (West and Fearnside, 2021). As a result, the high regional deforestation rates observed prior to 2004, used to inform the Brazilian project baselines, likely led to an overestimation of the projects' performance. Yet, unlike Brazil, the six countries in this study did not experience a similar level of reduction in deforestation nationwide after the REDD+ projects were implemented (Fig. S3). Hence, the seemingly unrealistic *ex-ante* baselines adopted by many projects likely resulted from the use of methodologies that systematically fail to produce credible counterfactuals for the REDD+ interventions, compromising the evaluation of the projects' performance at mitigating deforestation and thus climate change. This may be due to potentially three complementary reasons: poor foresight, oversight of temporal changes in deforestation drivers, and "gaming." First, projects may have unintentionally overestimated future deforestation pressures based on alarming historical trends that poorly represent current conditions. Second, baselines of voluntary REDD+ projects tend to be fixed for a period of 10 years, discouraging potential adjustments to reflect changes in deforestation drivers over time. Finally, baselines may have been strategically inflated to maximize revenues from offset sales.

In contrast, both standard and generalized SC methods use pre-REDD+ information to identify control areas, but use contemporaneous information on deforestation in the project and control areas to estimate additionality. Following well-accepted guidelines for rigorous impact evaluation (Ferraro and Hanauer, 2014), if properly selected, such *ex-post* counterfactuals can capture the effects of contemporaneous changes in deforestation drivers, thus being less biased by confounding external factors (Sills et al., 2017; West et al., 2020). If REDD+ projects were to adopt a similar dynamic approach, it would likely reduce the additionality problems with project baselines and offsets identified in this study. Promisingly, new methodological guidelines for future voluntary REDD+ projects may include the use of such "dynamic baselines" (Verra, 2021b).

Still, despite the clear advantages of using dynamic methods like the SC to construct *ex-post* deforestation baselines for REDD+ interventions, some implementation and monitoring challenges would likely arise from their adoption. First, given biophysical heterogeneity across tropical regions, ideal control areas for the project sites may not always be identified (Tables A1–A29), or they could be manipulated (e.g., intentionally degraded) to misleadingly improve project performance. Second, such dynamic baselines may still fail to account for all relevant determinants of deforestation due to data constraints. Finally, long-lasting voluntary REDD+ projects may eventually outlive suitable control sites needed to feed the dynamic baselines.

One alternative would be to require projects to adopt *ex-ante* jurisdictional baselines instead, pre-established by government agencies. While these baselines might still fall short of capturing contemporaneous changes in deforestation drivers, they could be updated more frequently than the individual project baselines so as to update recent deforestation pressures and spatial patterns. Most importantly, jurisdictional baselines may also better capture government efforts to control forest loss, thus mitigating the risk of wrongly attributing reductions in deforestation achieved through public policies to private REDD+ interventions (Sills et al., 2017). Transferring the responsibility of baseline construction from project developers to jurisdictions could also reduce the room for "baseline gaming." Hence, nesting voluntary projects into subnational jurisdictions appears to be a promising future pathway for REDD+, a practice being increasingly promoted worldwide (Lee et al., 2018).

But, turning to conservation performance, why have some projects seemingly failed to reduce deforestation at all? Some may have struggled with on-the-ground implementation and execution of envisioned conservation activities; others may have promoted ineffective actions—perhaps due to funding uncertainties, slow commercialization of carbon credits, or lack of experience (Laing et al., 2016; Seyller et al., 2016; Verra, 2018; Wunder et al., 2020). Notably, many projects claimed to have started much earlier than the year they were certified. While this allows projects to issue retroactive offsets right after certification (Linacre et al., 2015), it also implies that they may not have had access to funding during their initial years, potentially compromising the execution of planned conservation actions.

A recent evaluation on the effectiveness of the same type of REDD+ interventions reported significant reductions in deforestation rates, on average (Guizar-Coutiño et al., 2022). The study, based on gridded data, estimated an average deforestation rate of 0.2% year$^{-1}$ in the REDD+ sites versus 0.4% year$^{-1}$ for their matched control areas. The size of these estimates is in line with our results for the Peruvian and African projects, but in our case, they were statistically insignificant, potentially due to our lower sample sizes compared to the pixel-based samples from Guizar-

Coutiño et al. (2022). More importantly, from the offset perspective, the estimated average reductions—significant or not—remain substantially lower than reductions in deforestation claimed by the projects.

Our study provides further evidence on the effectiveness of voluntary REDD+ projects and questions their *de facto* additionality (Angelsen, 2008). Only a minority of the projects significantly reduced deforestation compared to the *ex-post* counterfactuals, and even those did not reduce deforestation to the extent claimed. Given that REDD+ payments are largely performance-based, and supported by those interested in offsetting their own emissions, only the offsets associated with additional reductions in deforestation should be eligible for trading in voluntary carbon markets (Angelsen, 2017; Seyller et al., 2016; Taskforce on Scaling Voluntary Carbon Markets, 2021). Certification schemes are allegedly in place to safeguard the additionality of offsets, but our results indicate that this is not enough. It is critical to develop new and rigorous methods for the construction of credible deforestation baselines for voluntary REDD+ interventions, and to properly and regularly assess their contribution to climate change mitigation.

Finally, the evidence from this and other studies indicate that some voluntary projects have effectively reduced deforestation (Guizar-Coutiño et al., 2022), particularly in Peru. For REDD+ to be scaled and achieve its ambitious goals worldwide, it is paramount that we better understand the factors responsible for their successes and failures, including their impacts on local communities. Researchers and practitioners will need to form effective partnerships to jointly meet these challenges, putting also a more realistic bar for forest carbon offsets that would help REDD+ initiatives fulfill their original promise.

## Materials and Methods

### Project sample

There are 27 voluntary avoided (unplanned) deforestation REDD+ projects in Peru, Colombia, DRC, Tanzania, Zambia, and Cambodia that were certified under Verra's *Verified Carbon Standard* (VCS; Verra, 2019) by July 2021 (Table 1; Fig. 2 & S1). We chose these six countries because they have a large number of projects and provide coverage of all tropical continents. We adopted VCS' unique IDs to identify the projects in our study. Because Project 1360 from Peru is composed of three disconnected areas, each of these areas was evaluated independently (identified with IDs 1360-1, 1360-2, and 1360-3). We adopted the same approach for Project 1775 from Tanzania. Project sites were defined by the geospatial polygons (i.e., KML files) provided by the project developers and available from the VCS project database (https://registry.verra.org/). While we intended to examine all VCS-certified projects in our focal countries, we removed projects that did not provide the correct geospatial data on their boundaries and those with corrupted KML files (i.e., Projects #868 from Peru and #1399 and #1695 from Colombia). Overall, officially reported project areas matched the areas of the KML files reasonably well, except for Project #1566 from Colombia and #1325 from Tanzania (Table S1).

Deforestation baselines adopted by these projects followed VCS-approved carbon-accounting methodologies (Table S1). While there are differences among these methodologies, they generally require baselines to be established *ex-ante* based on historical regional deforestation averages over 10-year intervals prior to project implementation.

**Individual synthetic control analyses**

Impact evaluations that define causality based on potential outcomes rely on the establishment of credible counterfactuals quantifying what would have happened in the absence of an intervention. These counterfactuals are inherently unobservable. The most common strategy for establishing robust counterfactuals is to observe outcomes in control areas, which are not exposed to the treatment under evaluation but which are otherwise very similar to those areas (Ferraro and Hanauer, 2014). Following West et al. (2020), we employed the SC method to construct project-specific weighted combinations of control areas, or "SCs," as our counterfactuals (Sills et al., 2020; Xu, 2017). We adopted this method due to the small sample size and likely heterogeneous effects of voluntary REDD+ projects (Abadie et al., 2021). SCs were constructed as a weighted average of control areas through a nested optimization procedure that minimizes the differences in pre-REDD+ characteristics of the project sites and their respective SCs (see Abadie et al., 2011, for details); the characteristics of the control areas (i.e., model covariates; Table S10) were weighted such that the resulting weighted average outcome of the selected areas most closely matched the cumulative pre-REDD+ deforestation rates in the project sites since 2001.

One of the key challenges with counterfactual-based impact evaluation is defining and characterizing potential controls, in this case, areas that could have been but are not REDD+ projects. West et al. (2020) defined potential controls as landholding polygons obtained from Brazil's national Rural Environmental Registry (CAR; Portuguese acronym), a spatially explicit database created to determine forest restoration requirements. However, similar databases are not available for the countries considered in this study. We addressed this limitation by creating a pool of 1000 circular spatial polygons, or "pseudo" control areas, for each project, randomly distributed across the focal countries, each the same size as the project site (Fig. S12). Prior to the construction of the SCs, we restricted our pool of control areas to a subset of potential SC "donors" that shared similar characteristics to the project sites. The inclusion of donors in the control sets was based on deforestation pressure (i.e., the average annual deforestation in the projects' 10-km buffer zones prior to the project start date). We first attempted to include donors with ±10% buffer deforestation as the project sites. We increased this range by an additional ±10% each time the resulting SC failed to replicate the historical deforestation pattern in the project site. This approach was of critical importance because, unlike in West et al. (2020), many of the projects' KML files from our sample were restricted to the forested areas within the project site at the start of project. Hence, without explicitly controlling for deforestation pressure surrounding the KML boundaries, the SCs would likely be based on control areas exposed to much lower risk of forest loss, rendering them poor counterfactuals for REDD+ sites (Guizar-Coutiño et al., 2022). Overall, most SCs experienced similar levels of buffer deforestation as the REDD+ sites (see Annex A for the covariate balance between projects and SCs).

To construct the synthetic controls, we used covariates (listed in Table S8) that have been found related to deforestation (Busch and Ferretti-Gallon, 2017). We also included pre-REDD+ (13) annual and (14) cumulative deforestation rates for the construction of the SCs. Annual deforestation data for the focal countries, from 2001 to 2020, were processed in Google Earth Engine based on the Global Forest Change (GFC) product (Hansen et al., 2013). Many remote sensing studies highlight the differences in deforestation rates between GFC and the numbers officially recognized by governments (Griffiths et al., 2018; Milodowski et al., 2017; Qin et al., 2019). Such differences emerge from different mapping methodologies and definitions of deforestation and forest degradation.

The individual SC analyses were conducted with the *Synth* package (v.1.1-5; Abadie et al., 2011) available for R software (v.4.1.0). Results from the nested optimizations were refined based on the augmented method proposed by Becker and Klößner (2018).

### Project-specific synthetic control validation

We validated the SC method following West et al. (2020), by constructing a SC for each project site based on data only from the first half of the pre-REDD+ period (i.e., training interval) and validated against the second half of the period (i.e., testing interval). In theory, the validity of the method would be empirically proven if the future REDD+ sites and their respective SCs shared a similar deforestation trend in the testing interval. This validation approach focuses on each REDD+ site individually. We considered the SC method validated for the sites in which the gaps between the project and SC deforestation at the end of the validation interval were lower than 0.5% of the project area (Tables S3 & Fig. S2). This "proof of concept" differs from standard model validation practices, because the donor areas selected to be part of the SCs based on the first half of the pre-REDD+ periods are not necessarily the same areas selected when the full pre-REDD+ period was used for the optimization. Still, such an exercise arguably increases the credibility of SC analyses that pass this validity test.

### Placebo tests

We assessed the statistical significance of our individual findings with a series of placebo tests, in which we create SCs for all areas in the project-specific donor pool subsets and compute the difference in cumulative deforestation between each placebo and its SC in the years after that REDD+ project was implemented. Because placebo areas are not exposed to REDD+, any differences in forest loss between placebos and their SCs are considered statistical "noise." Following the previous literature (Abadie et al., 2011; West et al., 2020), we discarded placebo tests with mean squared prediction error (MSPE) five times higher than the MSPE of its respective REDD+ project. We then used the gaps in deforestation between the remaining placebos and their respective SCs to create 95% confidence intervals around the mean placebo effect (generally close to zero).

## Generalized synthetic controls

Standard applications of the SC method target individual interventions (Abadie et al., 2021; Sills et al., 2015), which limits the generalization of results. The method developed by Xu (2017) proposes a generalization of the SC method, unified with a special case of the difference-in-differences estimator for causal inference in time-series cross-sectional data that relaxes the parallel trends assumption. This method is known as the *generalized synthetic control* (GSC). First, an interactive fixed-effects model is estimated based solely on the control group data, and a fixed number of *latent factors* (i.e., time-varying intercepts) is selected. Then, *factor loadings* (i.e., unit-specific intercepts) are estimated for each REDD+ project based on the linear projection of pre-REDD+ deforestation rates. This formulation allows for the estimation of the average treatment effect on the treated unit (ATT), which in our case, corresponds to the average relative REDD+ impact on deforestation (% year-1) across the project sites within a country or region. Last, the estimated latent factors and loadings are used to create GSCs, which in our case are the projects' average deforestation counterfactuals.

We adopted the GSC method as a way to generalize and estimate the average REDD+ project effect on deforestation in Peru, Colombia, and Africa (by analyzing the projects from DRC, Tanzania, and Zambia together because of their limited number of projects). In addition, because the GSCs are independent of the individual SCs—and based on annual, rather than cumulative, deforestation rates—they can also serve as robustness checks of the individual SC results. Following Xu (2017), we adopted Equation 1 as the underlying interactive two-way fixed-effects model of our GSC analyses:

$$D_{it} = \alpha_i + \xi_t + \delta_{it}T_{it} + X_{it}\beta + \lambda_i f_t + \varepsilon_{it} \qquad (1)$$

where, $D_{it}$ is the annual deforestation in project site $i$ in year $t$; $\alpha_i$ represents project site individual effects; $\xi_t$ captures time effects; $T_{it}$ is a dummy variable indicating project implementation; $\delta_{it}$ captures the heterogeneous REDD+ effect on unit $i$ at year $t$; $X_{it}$ is a matrix of observed time-variant covariates, i.e., precipitation and deforestation in 1-km and 10-km buffer zones (results from augmented Dickey-Fuller tests suggest stationarity of the deforestation time series at the unit level); $\beta$ is a vector of unknown parameters; $\lambda_i$ is a vector of unknown factor loadings; $f_t$ is a vector of unobserved common latent factors; and $\varepsilon_{it}$ represents unobserved idiosyncratic shocks with zero mean. Formulation 1 implies that time-invariant covariates are dropped from the model by demeaning. To improve the robustness of the GSC analyses, by indirectly considering the relevance of time-invariant covariates, the GSCs were constructed exclusively based on the same donor areas used to create the individual SCs.

The GSC analyses were implemented with the *gsynth* package (v.1.2.1) available for R software. The package allows for several different specifications. Estimates were produced with the Expectation Maximization algorithm (Gobillon and Magnac, 2016), which benefits from the project area information in the pre-REDD+ period. Due to the relatively low number of observations in our data, we adopted the *gsynth*'s matrix completion method as the GSC estimator (Athey et al., 2017). A cross-validation procedure was implemented to select the optimal number of factors (or "hyper-parameters") in the matrix completion algorithm, ranging from zero to five. Last, uncertainty estimates were produced based on 1000-bootstrap runs (see Xu, 2017, for details).

The GSC analyses were based on two independent sets of selected controls, one exclusively based on the donors selected for the construction of the individual SCs and the other based on control areas selected with the genetic matching technique (Alexis and Sekhin, 2013), independent of the SC analyses. Each project site was matched to 10 control areas. The genetic matching was performed with the *Matching* package (v.4.10-8) available for available for R software (Sekhon, 2011), based on pre-project covariate information described in Table S11 (Fig. S13–S15).

**Additional robustness test**

An additional robustness test was performed where operational REDD+ sites were matched with "to-be" REDD+ sites based on the matching-based methods for cross-sectional panel data analysis developed by Imai et al. (2018) and implemented with the *PanelMatch* package (v.1.0.0) available for R software (Kim et al., 2020). Specifically, each operational REDD+ site was matched to up to 10 not-yet-operational REDD+ sites (or a weighted combination of those) sharing similar covariate history over a 5-year interval prior to project implementation and that remained not-operational over the given evaluation period (Fig. S16 & S17). Matching was performed based on three methods: propensity score matching, Mahalanobis distance, and propensity score weighting (Fig. S18). Once the matching sets were constructed from each matching approach, ATTs were annually estimated for evaluation periods ranging 1–5 years following project implementation.

**Additionality of the carbon credits**

We estimated the volume of (additional) carbon offsets from the voluntary REDD+ projects as compared to a counterfactual based on the observed deforestation from the individual SCs. Credits from these projects are generally issued after third-party audits. These credits are based on the estimated carbon emission reductions from the avoided deforestation brought about by the project activities, calculated as the net difference between the carbon emissions under the baseline and the REDD+ scenario (West et al., 2020). Under this crediting system, the *ex-post* volume of carbon offsets generated by the REDD+ projects is determined by deforestation in the REDD+ project as compared to the *ex-ante* baseline. We adopted a simplified approach to estimate such volumes by assuming a linear, per-hectare relationship between the baseline deforestation adopted by projects and their reported *ex-ante* volume of credits to be generated through 2020. Thus, projects with insufficient public information about *ex-ante* annual baseline deforestation rates and carbon stock were excluded from this assessment rates (Table S10; Fig. 2 & S8).

First, we identified the projects with significantly lower deforestation rates than their SCs according to the placebo tests. Projects that failed to achieve significant reductions in deforestation were assumed not to have reduced net carbon emissions. Then, we estimated the volume of credits generated by each project that significantly reduced deforestation based on the difference in observed deforestation (ha) between the SC and the project sites. Finally, we compared our carbon offset estimates based on the SCs to the *ex-ante* volume of credits expected by the projects from the year of project implementation through 2020 (Table S10; Fig. 2 & S8).

## Acknowledgments

This research was supported by Norway's International Climate and Forest Initiative (NICFI), the Meridian Institute, the Center for International Forestry Research (CIFOR)'s Global Comparative Study on REDD+, the European Forest Institute's BMEL-financed NewGo project, and the German Development Agency (GIZ).

**Supplementary Information**

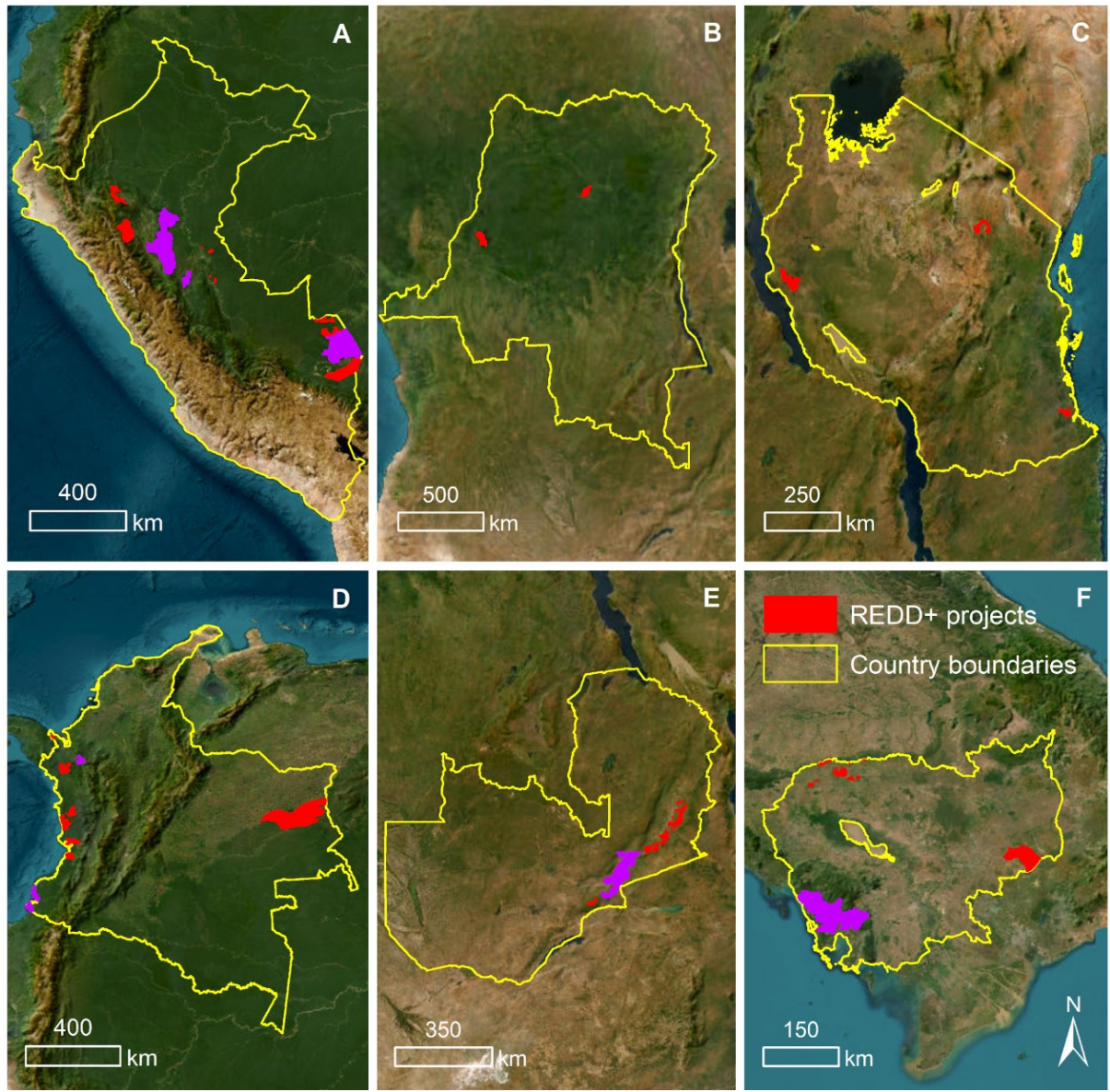

Figure S1. VCS-certified REDD+ project sites based on avoided unplanned deforestation implemented during 2009–2020 in Peru (A), Democratic Republic of Congo (B), Tanzania (C), Colombia (D), Zambia (E), and Cambodia (F). Purple areas are the sites excluded from the analysis.

Table S1. VCS-certified REDD+ projects based on avoided unplanned deforestation and degradation.

| Country | Project ID | Project name | Start year | Adopted VCS methodology | Included in the analysis | Reported project area (ha) | Project polygon area (ha) | Area overlap | Project polygon average tree cover in 2000* |
|---|---|---|---|---|---|---|---|---|---|
| Peru | 1882 | REDD+ Project in the Alto Huayabamba Conservation Concession (CCAH) | 2013 | VM0015 | Yes | 53,410 | 53,357 | 100% | 85% |
| | 1360 | Forest Management to reduce deforestation and degradation in Shipibo Conibo and Cacataibo indigenous communities of Ucayali region | 2010 | VM0015 | Partially‡ | 127,004 | 127,098 | 100% | 99% |
| | 2278 | The Jaguar Amazon REDD Project | 2018 | VM0006 | Yes | 183,015 | 183,120 | 100% | 99% |
| | 1067 | Reduction of deforestation and degradation in Tambopata National Reserve and Bahuaja-Sonene National Park within the area of Madre de Dios region–Peru | 2011 | VM0007 | Yes | 541,620 | 557,250 | 97% | 98% |
| | 985 | Cordillera Azul National Park REDD Project | 2009 | VM0007 | No† | 1,351,964 | 1,352,298 | 100% | 97% |
| | 958 | Biocorredor Martin Sagrado | 2011 | VM0015 | Yes | 295,654 | 295,412 | 100% | 93% |
| | 944 | Alto Mayo Conservation Initiative | 2009 | VM0015 | Yes | 182,000 | 177,533 | 103% | 89% |
| | 844 | Madre De Dios Amazon Redd Project | 2009 | VM0007 | Yes | 98,932 | 97,998 | 101% | 99% |
| Colombia | 1400 | Concosta REDD+ Project | 2013 | VM0006 | Yes | 54,623 | 63,961 | 85% | 97% |
| | 1566 | REDD+ Project Resguardo Indígena Unificado de la Selva de Matavén (RIU-SM) | 2013 | VM0007 | Yes | 1,150,212 | 1,753,035 | 66% | 85% |
| | 856 | The Chocó-Darién Conservation Corridor REDD Project | 2011 | VM0009 | Yes | 13,465 | 12,710 | 106% | 95% |
| | 1396 | Rio Pepe y ACABA REDD+ Project | 2014 | VM0006 | Yes | 48,177 | 57,051 | 84% | 98% |
| | 1395 | Bajo Calima y Bahía Málaga (BCBM) REDD+ Project | 2013 | VM0006 | Yes | 83,452 | 91,831 | 91% | 98% |
| | 1392 | Cajambre REDD+ Project | 2013 | VM0006 | Yes | 661,69 | 60,316 | 110% | 98% |
| | 1391 | Sivirú, Usaragá, Pizarro y Pilizá (SUPP) REDD+ Project | 2013 | VM0006 | Yes | 47,667 | 55,104 | 87% | 98% |
| | 1390 | Carmen del Darién (CDD) REDD+ Project | 2014 | VM0006 | Yes | 118,318 | 131,828 | 90% | 94% |
| | 1389 | ACAPA – Bajo Mira y Frontera (ACAPA-BMF) REDD+ Project | 2013 | VM0006 | No† | 58,212 | 68,602 | 85% | 95% |

Table S1 (continued). VCS-certified REDD+ projects based on avoided unplanned deforestation and degradation.

| Country | Project ID | Project name | Start year | Adopted VCS methodology | Included in the analysis | Reported project area (ha) | Project polygon area (ha) | Area overlap | Project polygon average tree cover in 2000* |
|---|---|---|---|---|---|---|---|---|---|
| Cambodia | 1748 | Southern Cardamom REDD+ Project | 2015 | VM0009 | No† | 445,339 | 458,408 | 97% | 93% |
| Cambodia | 904 | Reduced Emissions from Deforestation and Degradation in Community Forests – Oddar Meanchey, Cambodia | 2008 | VM0006 | Yes | 56,050 | 66,205 | 85% | 42% |
| Cambodia | 1650 | Reduced Emissions from Deforestation and Degradation in Keo Seima Wildlife Sanctuary | 2010 | VM0015 | Yes | 166,983 | 193,503 | 86% | 73% |
| DRC | 934 | The Mai Ndombe REDD+ Project | 2011 | VM0009 | Yes | 299,640 | 301,263 | 99% | 89% |
| DRC | 1359 | Isangi REDD+ Project | 2009 | VM0006 | Yes | 187,571 | 188,489 | 100% | 100% |
| Tanzania | 1325 | Mjumita Community Forest Project (LINDI) | 2011 | VM0015 | Yes | 41,924 | 65,279 | 64% | 51% |
| Tanzania | 1900 | Makame Savannah REDD | 2016 | VM0007 | Yes | 104,065 | 107,152 | 97% | 10% |
| Tanzania | 1897 | Ntakata Mountains REDD | 2017 | VM0007 | Yes | 216,994 | 204,203 | 106% | 50% |
| Zambia | 1775 | Luangwa Community Forests Project | 2015 | VM0009 | Partially‡ | 943,676 | 943,674 | 100% | 23% |
| Zambia | 1202 | Lower Zambezi REDD+ Project | 2009 | VM0009 | Yes | 40,126 | 40,103 | 100% | 24% |

* Processed in Google Earth Engine based on the 2000 Percent Tree Cover map from the Global Forest Change dataset (Hansen et al., 2013).

† Discarded due to the poor quality of the synthetic control counterfactual.

‡ Project composed of multiple sites, some of which were discarded due to the poor quality of the synthetic control counterfactual.

Source: Verified Carbon Standard (VCS) project database (https://registry.verra.org/).

Table S2. VCS-certified REDD+ projects: carbon offsets issued and retired as of November 2021.

| Country | Project ID | Project name | Start year | Adopted VCS methodology | Carbon offsets issued | Carbon offsets retired | Retired proportion (%) |
|---|---|---|---|---|---|---|---|
| Peru | 1882 | REDD+ Project in the Alto Huayabamba Conservation Concession (CCAH) | 2013 | VM0015 | 171,673 | 14,443 | 8.4 |
| | 1360 | Forest Management to reduce deforestation and degradation in Shipibo Conibo and Cacataibo indigenous communities of Ucayali region | 2010 | VM0015 | 4,852,836 | 763,786 | 15.7 |
| | 2278 | The Jaguar Amazon REDD Project | 2018 | VM0006 | 0 | 0 | – |
| | 1067 | Reduction of deforestation and degradation in Tambopata National Reserve and Bahuaja-Sonene National Park within the area of Madre de Dios region–Peru | 2011 | VM0007 | 3,678,270 | 704,766 | 19.2 |
| | 985 | Cordillera Azul National Park REDD Project | 2009 | VM0007 | 25,240,371 | 6,323,967 | 25.1 |
| | 958 | Biocorredor Martin Sagrado | 2011 | VM0015 | 566,843 | 335,957 | 59.3 |
| | 944 | Alto Mayo Conservation Initiative | 2009 | VM0015 | 5,282,313 | 3,625,820 | 68.6 |
| | 844 | Madre De Dios Amazon Redd Project | 2009 | VM0007 | 9,658,069 | 975,539 | 10.1 |
| Colombia | 1400 | Concosta REDD+ Project | 2013 | VM0006 | 544,278 | 303,004 | 55.7 |
| | 1566 | REDD+ Project Resguardo Indígena Unificado de la Selva de Matavén (RIU-SM) | 2013 | VM0007 | 22,274,745 | 6,068,608 | 27.2 |
| | 856 | The Chocó-Darién Conservation Corridor REDD Project | 2011 | VM0009 | 435,188 | 158,822 | 36.5 |
| | 1396 | Rio Pepe y ACABA REDD+ Project | 2014 | VM0006 | 567,286 | 419,209 | 73.9 |
| | 1395 | Bajo Calima y Bahía Málaga (BCBM) REDD+ Project | 2013 | VM0006 | 1,620,202 | 747,709 | 46.1 |
| | 1392 | Cajambre REDD+ Project | 2013 | VM0006 | 477,432 | 290,483 | 60.8 |
| | 1391 | Sivirú, Usaragá, Pizarro y Pilizá (SUPP) REDD+ Project | 2013 | VM0006 | 1,021,146 | 625,751 | 61.3 |
| | 1390 | Carmen del Darién (CDD) REDD+ Project | 2014 | VM0006 | 681,565 | 398,348 | 58.4 |
| | 1389 | ACAPA – Bajo Mira y Frontera (ACAPA-BMF) REDD+ Project | 2013 | VM0006 | 783,133 | 323,456 | 41.3 |

Table S2 (continued). VCS-certified REDD+ projects: carbon offsets issued and retired as of November 2021.

| Country | Project ID | Project name | Start year | Adopted VCS methodology | Carbon offsets issued | Carbon offsets retired | Retired proportion (%) |
|---|---|---|---|---|---|---|---|
| Cambodia | 1748 | Southern Cardamom REDD+ Project | 2015 | VM0009 | 23,785,965 | 2,365,202 | 9.9 |
| | 904 | Reduced Emissions from Deforestation and Degradation in Community Forests – Oddar Meanchey, Cambodia | 2008 | VM0006 | 48,000 | 44,297 | 92.3 |
| | 1650 | Reduced Emissions from Deforestation and Degradation in Keo Seima Wildlife Sanctuary | 2010 | VM0015 | 14,568,314 | 154,747 | 1.1 |
| DRC | 934 | The Mai Ndombe REDD+ Project | 2011 | VM0009 | 13,322,276 | 2,572,569 | 19.3 |
| | 1359 | Isangi REDD+ Project | 2009 | VM0006 | 1,620,202 | 747,709 | 46.1 |
| Tanzania | 1325 | Mjumita Community Forest Project (LINDI) | 2011 | VM0015 | 10,000 | 2246 | 22.5 |
| | 1900 | Makame Savannah REDD | 2016 | VM0007 | 150,425 | 80,000 | 53.2 |
| | 1897 | Ntakata Mountains REDD | 2017 | VM0007 | 726,000 | 84,115 | 11.6 |
| Zambia | 1775 | Luangwa Community Forests Project | 2015 | VM0009 | 6,147,649 | 1,911,839 | 31.1 |
| | 1202 | Lower Zambezi REDD+ Project | 2009 | VM0009 | 1,570,196 | 689,134 | 43.9 |
| TOTAL | – | – | – | – | 139,804,377 | 38,154,319 | 22.3 |

Source: Verified Carbon Standard (VCS) project database (https://registry.verra.org/).

***Standard synthetic control validation.*** Before assessing the impacts of the REDD+ projects, we explored whether the synthetic controls could accurately replicate deforestation trends in the project sites prior to project implementation. Synthetic controls were able to replicate pre-REDD+ deforestation trends reasonably well in most project sites (Table S3 & Fig. S2).

Table S3. Synthetic control validation: difference between project and synthetic control deforestation based on prior to project implementation.

| Country | Project ID | Final year of the validation period | Project deforestation (ha) | Synthetic control deforestation (ha) | Difference between project and synthetic control deforestation | |
|---|---|---|---|---|---|---|
| | | | | | (ha) | (% of project area) |
| Cambodia | 1650 | 2010 | 2389.8 | 2142.0 | 247.8 | 0.1 |
| Cambodia | 1748 | 2015 | 3089.9 | 2700.8 | 389.0 | 0.1 |
| Cambodia | 904 | 2007 | 1122.3 | 1032.2 | 90.0 | 0.1 |
| Colombia | 1389 | 2013 | 929.9 | 927.2 | 2.7 | 0.0 |
| Colombia | 1390 | 2014 | 365.1 | 565.6 | −200.5 | −0.2 |
| Colombia | 1391 | 2013 | 55.2 | 190.9 | −135.8 | −0.2 |
| Colombia | 1392 | 2013 | 132.0 | 89.0 | 43.0 | 0.1 |
| Colombia | 1395 | 2013 | 554.8 | 890.7 | −335.8 | −0.4 |
| Colombia | 1396 | 2014 | 643.3 | 659.2 | −16.0 | 0.0 |
| Colombia | 1400 | 2013 | 89.7 | 268.5 | −178.7 | −0.3 |
| Colombia | 1566 | 2013 | 10,761.7 | 9563.2 | 1198.5 | 0.1 |
| Colombia | 856 | 2011 | 88.6 | 147.9 | −59.4 | −0.5 |
| DRC | 1359 | 2008 | 1509.9 | 1572.4 | −62.4 | 0.0 |
| DRC | 934 | 2010 | 8234.9 | 7975.1 | 259.8 | 0.1 |
| Peru | 1067 | 2011 | 1953.1 | 1687.8 | 265.4 | 0.0 |
| Peru | 1182 | 2013 | 243.6 | 252.3 | −8.8 | 0.0 |
| Peru | 1360-1 | 2010 | 381.2 | 364.8 | 16.3 | 0.0 |
| Peru | 1360-2 | 2010 | 34.7 | 31.6 | 3.0 | 0.0 |
| Peru | 1360-3 | 2010 | 73.2 | 73.8 | −0.5 | 0.0 |
| Peru | 2278 | 2018 | 995.8 | 718.3 | 277.4 | 0.2 |
| Peru | 844 | 2009 | 50.8 | 60.5 | −9.7 | 0.0 |
| Peru | 944 | 2009 | 2626.6 | 2843.2 | −216.6 | −0.1 |
| Peru | 958 | 2011 | 1504.6 | 1749.0 | −244.4 | −0.1 |
| Peru | 985 | 2009 | 2914.3 | 2224.3 | 690.0 | 0.1 |
| Tanzania | 1325 | 2011 | 2913.9 | 3456.2 | −542.3 | −0.8* |
| Tanzania | 1897 | 2017 | 4279.0 | 7701.1 | −3422.2 | −1.7* |
| Tanzania | 1900 | 2016 | 16.3 | 7.1 | 9.2 | 0.0 |
| Zambia | 1202 | 2009 | 193.5 | 37.7 | 155.8 | 0.4 |
| Zambia | 1775-1 | 2015 | 1526.2 | 779.7 | 746.6 | 0.1 |
| Zambia | 1775-2 | 2015 | 136.5 | 406.9 | −270.4 | −0.2 |
| Zambia | 1775-3 | 2015 | 197.2 | 157.8 | 39.4 | 0.0 |

*Projects with difference between project and synthetic control deforestation >0.5% of the project area are assumed to have failed validation.

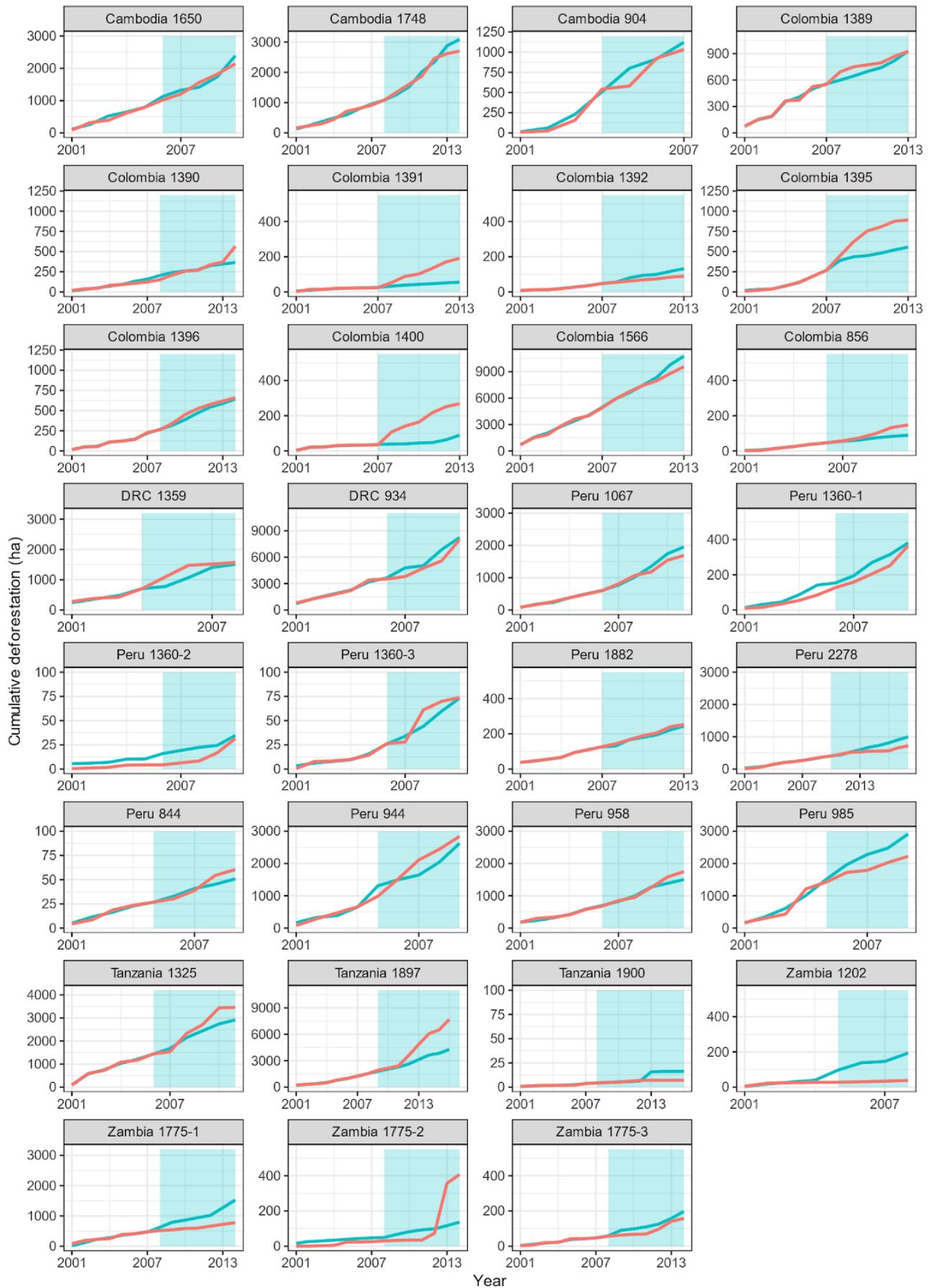

Figure S2. Validation of the synthetic control method. Pre-REDD+ deforestation in "to-be" REDD+ project areas (red) versus synthetic controls (blue). Shaded blue areas represent the validation periods (note: scales differ).

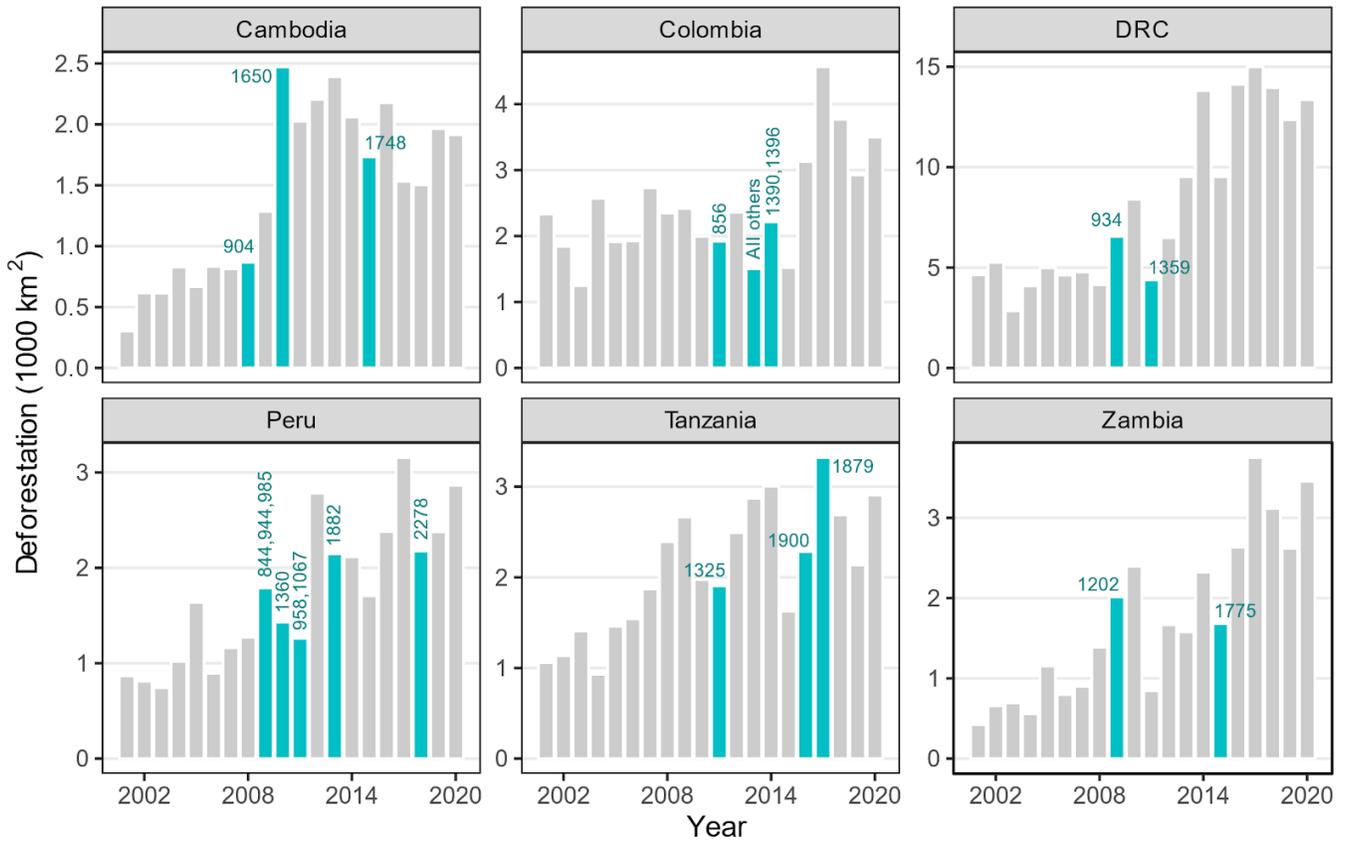

Figure S3. Annual deforestation rates (1000 km²) in the study countries based on the Global Forest Change dataset (bars). Blue bars indicate the implementation year of the VCS-certified REDD+ projects. Project IDs displayed in blue.

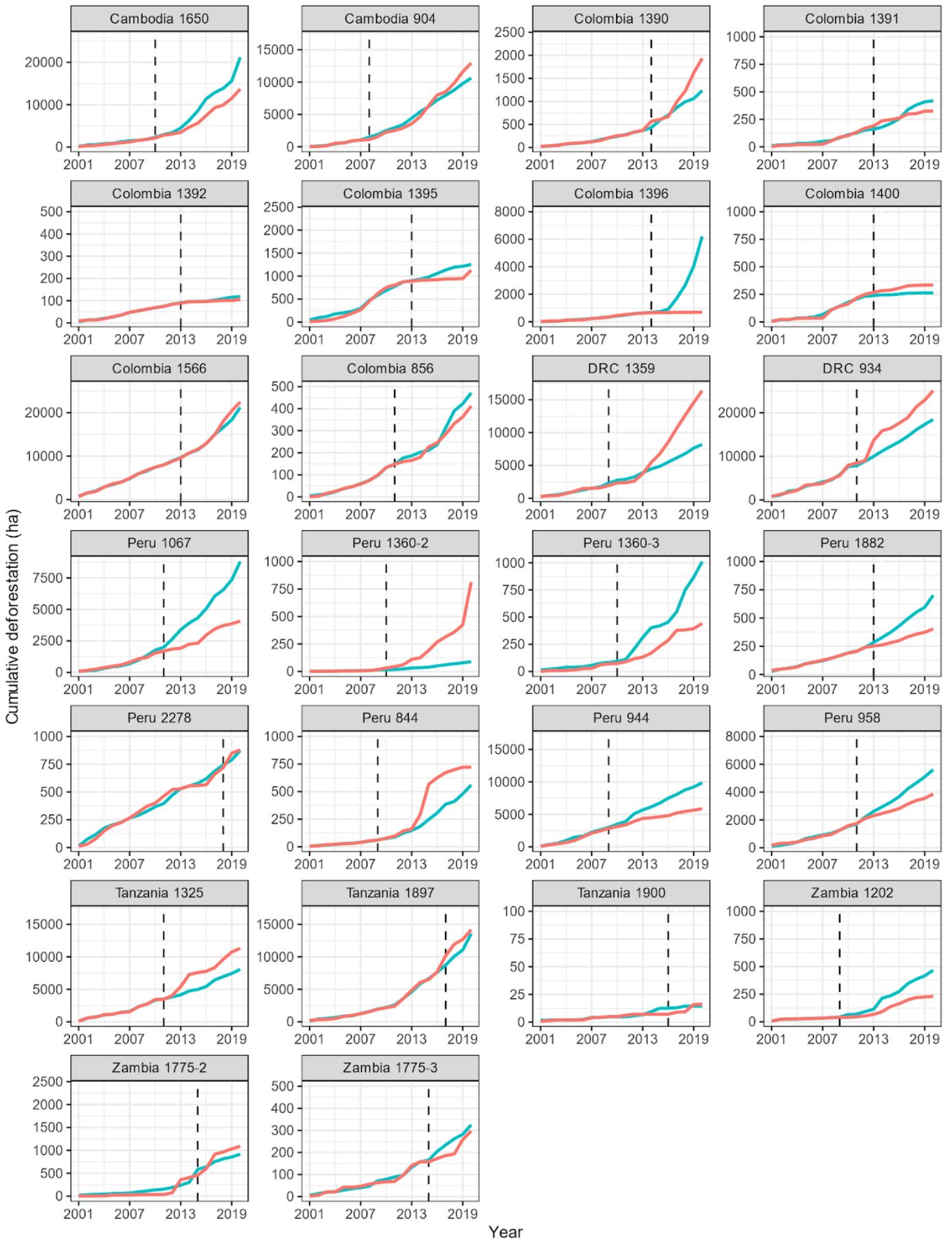

Figure S4. Cumulative post-2000 deforestation in REDD+ project areas (red) versus synthetic controls (blue). Dashed black lines indicate the project implementation year (note: scales differ).

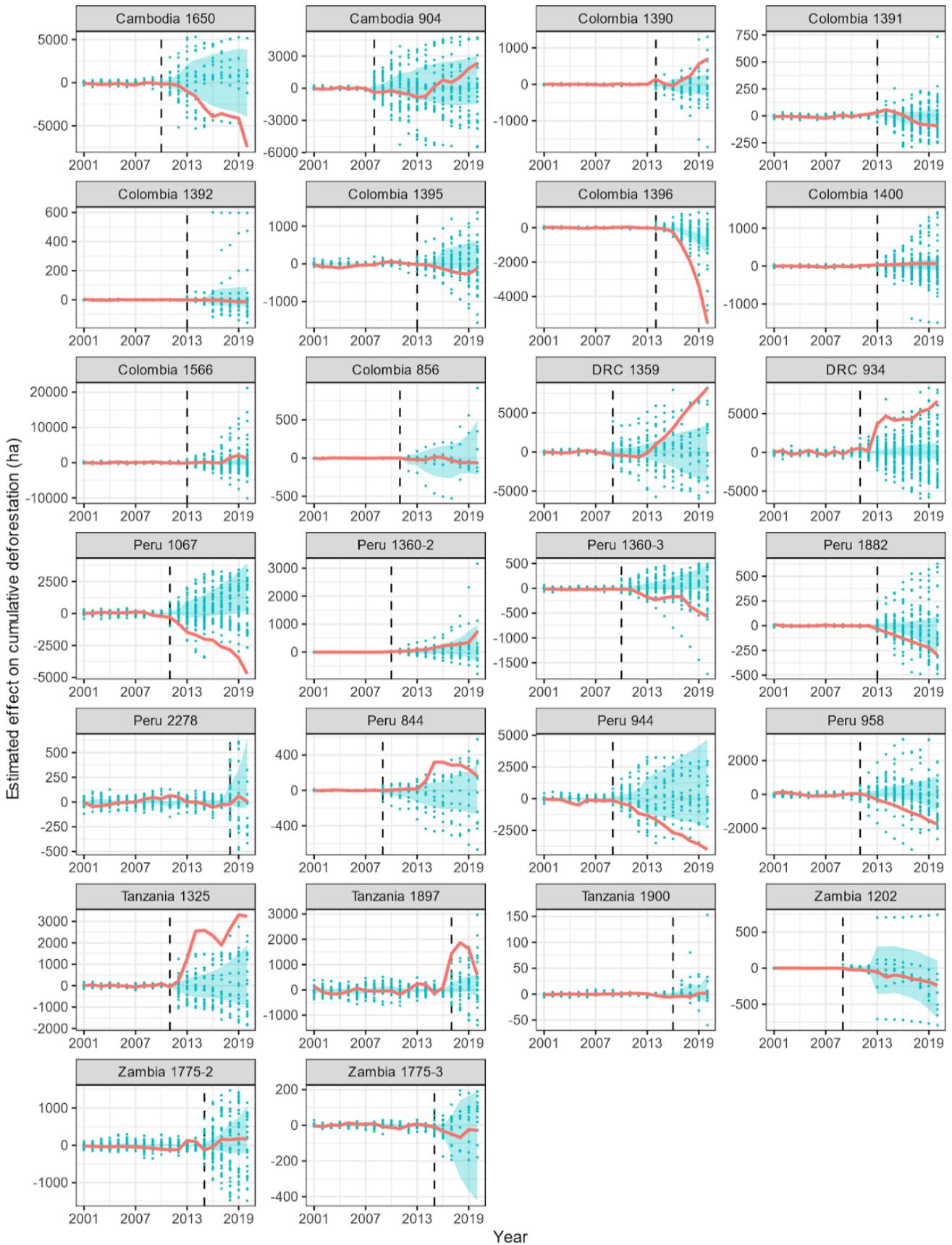

Figure S5. Placebo tests: cumulative deforestation in REDD+ project areas minus deforestation in their respective synthetic controls (red) and placebos minus their respective synthetic controls (blue dots). Dashed black lines indicate the project implementation year (assumed the same for placebos). Shaded blue areas represent 95% confidence intervals around the placebos' mean (note: scales differ).

Table S4. Estimated average REDD+ effect of the ***Peruvian projects*** on deforestation based on the selected control areas from the <u>individual synthetic controls</u>.

| Years relative to project implementation | Average REDD+ effect on forest loss (ATT*; %) | Standard error | Confidence interval | | p-value | Sample size (projects) |
|---|---|---|---|---|---|---|
| | | | Lower | Upper | | |
| −7 | 0.0568 | 0.0183 | 0.0209 | 0.0927 | 0.0019 | 0 |
| −6 | 0.0178 | 0.0138 | −0.0093 | 0.0449 | 0.1969 | 0 |
| −5 | 0.0407 | 0.0193 | 0.0029 | 0.0785 | 0.0348 | 0 |
| −4 | 0.0104 | 0.0219 | −0.0324 | 0.0533 | 0.6335 | 0 |
| −3 | −0.0258 | 0.0263 | −0.0774 | 0.0258 | 0.3267 | 0 |
| −2 | 0.0026 | 0.0223 | −0.0411 | 0.0462 | 0.9077 | 0 |
| −1 | −0.0339 | 0.0251 | −0.0831 | 0.0153 | 0.1769 | 0 |
| 0 | −0.0974 | 0.0259 | −0.1482 | −0.0466 | 2.00E−04 | 0 |
| 1 | −0.1647 | 0.0659 | −0.2939 | −0.0355 | 0.0125 | 10 |
| 2 | −0.1884 | 0.0887 | −0.3623 | −0.0145 | 0.0337 | 10 |
| 3 | −0.2209 | 0.0756 | −0.3691 | −0.0727 | 0.0035 | 10 |
| 4 | −0.2643 | 0.1499 | −0.5582 | 0.0296 | 0.078 | 9 |
| 5 | −0.236 | 0.152 | −0.5339 | 0.062 | 0.1206 | 9 |
| 6 | −0.1472 | 0.1163 | −0.3751 | 0.0807 | 0.2055 | 9 |
| 7 | −0.1193 | 0.1369 | −0.3876 | 0.1491 | 0.3837 | 9 |
| 8 | −0.3545 | 0.17 | −0.6877 | −0.0212 | 0.0371 | 9 |
| 9 | −0.2853 | 0.2092 | −0.6953 | 0.1247 | 0.1726 | 8 |
| 10 | −0.2817 | 0.2581 | −0.7875 | 0.2241 | 0.2751 | 8 |
| 11 | 0.01 | 0.2993 | −0.5765 | 0.5965 | 0.9734 | 6 |
| 12 | −0.7423 | 0.4433 | −1.6111 | 0.1265 | 0.094 | 3 |
| Mean ATT* | −0.2253 | 0.1362 | −0.4923 | 0.0416 | 0.0981 | – |

*Average treatment effect on the treated.

Table S5. Estimated average REDD+ effect of the *Colombian projects* on deforestation based on the selected control areas from the individual synthetic controls.

| Years relative to project implementation | Average REDD+ effect on forest loss (ATT*; %) | Standard error | Confidence interval | | p-value | Sample size (projects) |
|---|---|---|---|---|---|---|
| | | | Lower | Upper | | |
| −9 | 0.0108 | 0.0106 | −0.0099 | 0.0315 | 0.3074 | 0 |
| −8 | −0.0111 | 0.0127 | −0.0361 | 0.0138 | 0.3819 | 0 |
| −7 | 0.0304 | 0.013 | 0.0048 | 0.056 | 0.0198 | 0 |
| −6 | −0.0284 | 0.0148 | −0.0574 | 5.00E−04 | 0.0545 | 0 |
| −5 | −4.00E−04 | 0.015 | −0.0297 | 0.029 | 0.9813 | 0 |
| −4 | 0.0144 | 0.0166 | −0.0182 | 0.047 | 0.3862 | 0 |
| −3 | −0.0175 | 0.0163 | −0.0496 | 0.0145 | 0.2832 | 0 |
| −2 | −0.0064 | 0.0147 | −0.0351 | 0.0224 | 0.6647 | 0 |
| −1 | 0.0153 | 0.0183 | −0.0206 | 0.0511 | 0.4046 | 0 |
| 0 | 0.0062 | 0.0121 | −0.0176 | 0.0299 | 0.6106 | 0 |
| 1 | 0.0399 | 0.0263 | −0.0116 | 0.0914 | 0.1287 | 9 |
| 2 | 0.0348 | 0.0297 | −0.0235 | 0.0931 | 0.2416 | 9 |
| 3 | 0.0224 | 0.0335 | −0.0432 | 0.088 | 0.5037 | 9 |
| 4 | −0.0366 | 0.0605 | −0.1552 | 0.082 | 0.5454 | 9 |
| 5 | −0.0474 | 0.0476 | −0.1407 | 0.0459 | 0.3196 | 9 |
| 6 | −0.1223 | 0.0741 | −0.2676 | 0.0229 | 0.0989 | 9 |
| 7 | −0.0537 | 0.0672 | −0.1854 | 0.0779 | 0.4237 | 9 |
| 8 | −0.0256 | 0.1028 | −0.227 | 0.1758 | 0.803 | 7 |
| 9 | 0.0503 | 0.099 | −0.1437 | 0.2442 | 0.6113 | 1 |
| 10 | 0.1924 | 0.1496 | −0.1007 | 0.4856 | 0.1982 | 1 |
| Mean ATT* | −0.0195 | 0.0384 | −0.0948 | 0.0558 | 0.6118 | – |

*Average treatment effect on the treated.

Table S6. Estimated average REDD+ effect of the *African projects* on deforestation based on the selected control areas from the individual synthetic controls.

| Years relative to project implementation | Average REDD+ effect on forest loss (ATT*; %) | Standard error | Confidence interval | | p-value | Sample size (projects) |
|---|---|---|---|---|---|---|
| | | | Lower | Upper | | |
| −7 | −0.0375 | 0.0213 | −0.0793 | 0.0043 | 0.0784 | 0 |
| −6 | −0.0105 | 0.017 | −0.0437 | 0.0228 | 0.5371 | 0 |
| −5 | 0.021 | 0.027 | −0.0318 | 0.0739 | 0.4359 | 0 |
| −4 | −0.0294 | 0.02 | −0.0685 | 0.0098 | 0.1415 | 0 |
| −3 | −0.0409 | 0.0182 | −0.0765 | −0.0053 | 0.0244 | 0 |
| −2 | 0.0042 | 0.0208 | −0.0365 | 0.0449 | 0.8387 | 0 |
| −1 | 0.0098 | 0.0155 | −0.0205 | 0.0401 | 0.526 | 0 |
| 0 | 0.0234 | 0.0242 | −0.024 | 0.0709 | 0.333 | 0 |
| 1 | −0.1205 | 0.1065 | −0.3292 | 0.0882 | 0.2579 | 9 |
| 2 | −0.1319 | 0.0833 | −0.2952 | 0.0314 | 0.1134 | 9 |
| 3 | 0.2803 | 0.2113 | −0.1339 | 0.6945 | 0.1848 | 9 |
| 4 | 0.0664 | 0.1864 | −0.2991 | 0.4318 | 0.7219 | 9 |
| 5 | −0.5484 | 0.3778 | −1.2889 | 0.1921 | 0.1467 | 8 |
| 6 | −0.3244 | 0.2608 | −0.8356 | 0.1868 | 0.2136 | 7 |
| 7 | −0.2671 | 0.2169 | −0.6921 | 0.158 | 0.2182 | 4 |
| 8 | −0.4143 | 0.6538 | −1.6958 | 0.8672 | 0.5263 | 4 |
| 9 | −0.7475 | 0.6862 | −2.0924 | 0.5974 | 0.276 | 4 |
| 10 | −0.4781 | 0.5572 | −1.5702 | 0.6141 | 0.3909 | 4 |
| 11 | −0.0446 | 0.1506 | −0.3398 | 0.2507 | 0.7673 | 2 |
| 12 | −0.3831 | 0.2677 | −0.9078 | 0.1416 | 0.1525 | 2 |
| Mean ATT* | −0.2013 | 0.1777 | −0.5496 | 0.147 | 0.2573 | – |

*Average treatment effect on the treated.

Table S7. Estimated average REDD+ effect of the ***Peruvian projects*** on deforestation based on the selected control areas from the <u>genetic matching</u>.

| Years relative to project implementation | Average REDD+ effect on forest loss (ATT*; %) | Standard error | Confidence interval | | p-value | Sample size (projects) |
|---|---|---|---|---|---|---|
| | | | Lower | Upper | | |
| −7 | 0.0346 | 0.0180 | −0.0006 | 0.0698 | 0.0544 | 0 |
| −6 | −0.0093 | 0.0142 | −0.0371 | 0.0185 | 0.5118 | 0 |
| −5 | 0.0300 | 0.0144 | 0.0018 | 0.0583 | 0.0372 | 0 |
| −4 | 0.0173 | 0.0189 | −0.0197 | 0.0544 | 0.3595 | 0 |
| −3 | −0.0186 | 0.0215 | −0.0607 | 0.0236 | 0.3879 | 0 |
| −2 | −0.0009 | 0.0172 | −0.0346 | 0.0328 | 0.9595 | 0 |
| −1 | −0.0150 | 0.0204 | −0.0551 | 0.0250 | 0.4627 | 0 |
| 0 | −0.0616 | 0.0204 | −0.1015 | −0.0217 | 0.0025 | 0 |
| 1 | −0.2586 | 0.0997 | −0.4540 | −0.0631 | 0.0095 | 10 |
| 2 | −0.2161 | 0.1461 | −0.5024 | 0.0702 | 0.1390 | 10 |
| 3 | −0.3496 | 0.1097 | −0.5646 | −0.1347 | 0.0014 | 10 |
| 4 | −0.4890 | 0.1767 | −0.8352 | −0.1428 | 0.0056 | 9 |
| 5 | −0.4058 | 0.1954 | −0.7887 | −0.0228 | 0.0378 | 9 |
| 6 | −0.3026 | 0.1272 | −0.5519 | −0.0534 | 0.0173 | 9 |
| 7 | −0.3171 | 0.1480 | −0.6071 | −0.0271 | 0.0321 | 9 |
| 8 | −0.5807 | 0.1807 | −0.9348 | −0.2267 | 0.0013 | 9 |
| 9 | −0.5316 | 0.2265 | −0.9755 | −0.0877 | 0.0189 | 8 |
| 10 | −0.6137 | 0.2324 | −1.0691 | −0.1583 | 0.0083 | 8 |
| 11 | −0.3770 | 0.2641 | −0.8945 | 0.1406 | 0.1534 | 6 |
| 12 | −1.0598 | 0.5736 | −2.1840 | 0.0644 | 0.0646 | 3 |
| Mean ATT* | −0.4170 | 0.1467 | −0.7046 | −0.1294 | 0.0045 | – |

*Average treatment effect on the treated.

Table S8. Estimated average REDD+ effect of the *Colombian projects* on deforestation based on the selected control areas from the genetic matching.

| Years relative to project implementation | Average REDD+ effect on forest loss (ATT*; %) | Standard error | Confidence interval | | p-value | Sample size (projects) |
|---|---|---|---|---|---|---|
| | | | Lower | Upper | | |
| −9 | 0.0093 | 0.0142 | −0.0185 | 0.0371 | 0.5121 | 0 |
| −8 | −0.0148 | 0.0151 | −0.0444 | 0.0149 | 0.3286 | 0 |
| −7 | 0.0211 | 0.0156 | −0.0095 | 0.0516 | 0.1761 | 0 |
| −6 | −0.0211 | 0.0139 | −0.0484 | 0.0061 | 0.1289 | 0 |
| −5 | 0.0114 | 0.0168 | −0.0215 | 0.0442 | 0.4968 | 0 |
| −4 | −0.0008 | 0.0158 | −0.0317 | 0.0302 | 0.9612 | 0 |
| −3 | −0.0244 | 0.0248 | −0.0729 | 0.0241 | 0.3247 | 0 |
| −2 | −0.0077 | 0.0129 | −0.0331 | 0.0177 | 0.5518 | 0 |
| −1 | 0.0155 | 0.0210 | −0.0258 | 0.0567 | 0.4624 | 0 |
| 0 | 0.0014 | 0.0138 | −0.0256 | 0.0284 | 0.9169 | 0 |
| 1 | 0.0301 | 0.0325 | −0.0336 | 0.0938 | 0.3542 | 9 |
| 2 | −0.0020 | 0.0323 | −0.0653 | 0.0614 | 0.9519 | 9 |
| 3 | −0.0377 | 0.0519 | −0.1395 | 0.0641 | 0.4679 | 9 |
| 4 | −0.1353 | 0.0960 | −0.3234 | 0.0528 | 0.1587 | 9 |
| 5 | −0.0619 | 0.0573 | −0.1742 | 0.0504 | 0.2802 | 9 |
| 6 | −0.1273 | 0.0949 | −0.3133 | 0.0588 | 0.1800 | 9 |
| 7 | −0.0937 | 0.0725 | −0.2358 | 0.0484 | 0.1963 | 9 |
| 8 | −0.0507 | 0.1097 | −0.2656 | 0.1642 | 0.6439 | 7 |
| 9 | 0.0800 | 0.1131 | −0.1417 | 0.3018 | 0.4795 | 1 |
| 10 | 0.2448 | 0.1719 | −0.0921 | 0.5816 | 0.1544 | 1 |
| Mean ATT* | −0.0539 | 0.0568 | −0.1653 | 0.0575 | 0.3432 | – |

*Average treatment effect on the treated.

Table S9. Estimated average REDD+ effect of the *African projects* on deforestation based on the selected control areas from the genetic matching.

| Years relative to project implementation | Average REDD+ effect on forest loss (ATT*; %) | Standard error | Confidence interval | | p-value | Sample size (projects) |
|---|---|---|---|---|---|---|
| | | | Lower | Upper | | |
| −7 | −0.0294 | 0.0139 | −0.0565 | −0.0022 | 0.0340 | 0 |
| −6 | 0.0054 | 0.0177 | −0.0293 | 0.0400 | 0.7610 | 0 |
| −5 | −0.0078 | 0.0288 | −0.0642 | 0.0486 | 0.7866 | 0 |
| −4 | −0.0115 | 0.0159 | −0.0427 | 0.0197 | 0.4711 | 0 |
| −3 | −0.0492 | 0.0254 | −0.0991 | 0.0006 | 0.0529 | 0 |
| −2 | 0.0002 | 0.0123 | −0.0238 | 0.0242 | 0.9858 | 0 |
| −1 | 0.0242 | 0.0234 | −0.0217 | 0.0701 | 0.3007 | 0 |
| 0 | 0.0234 | 0.0246 | −0.0247 | 0.0715 | 0.3399 | 0 |
| 1 | −0.0645 | 0.0604 | −0.1830 | 0.0539 | 0.2856 | 9 |
| 2 | −0.0570 | 0.0425 | −0.1402 | 0.0262 | 0.1795 | 9 |
| 3 | 0.1536 | 0.1469 | −0.1343 | 0.4415 | 0.2958 | 9 |
| 4 | 0.0524 | 0.1797 | −0.2998 | 0.4046 | 0.7706 | 9 |
| 5 | −0.3026 | 0.2012 | −0.6969 | 0.0917 | 0.1326 | 8 |
| 6 | −0.2556 | 0.1935 | −0.6348 | 0.1237 | 0.1866 | 7 |
| 7 | −0.2515 | 0.1548 | −0.5550 | 0.0519 | 0.1043 | 4 |
| 8 | −0.2118 | 0.3748 | −0.9463 | 0.5227 | 0.5720 | 4 |
| 9 | −0.3399 | 0.3851 | −1.0948 | 0.4149 | 0.3775 | 4 |
| 10 | −0.2594 | 0.3383 | −0.9224 | 0.4037 | 0.4433 | 4 |
| 11 | −0.2117 | 0.2055 | −0.6144 | 0.1911 | 0.3030 | 2 |
| 12 | −0.3983 | 0.2571 | −0.9022 | 0.1057 | 0.1214 | 2 |
| Mean ATT* | −0.1256 | 0.1285 | −0.3774 | 0.1262 | 0.3281 | – |

*Average treatment effect on the treated.

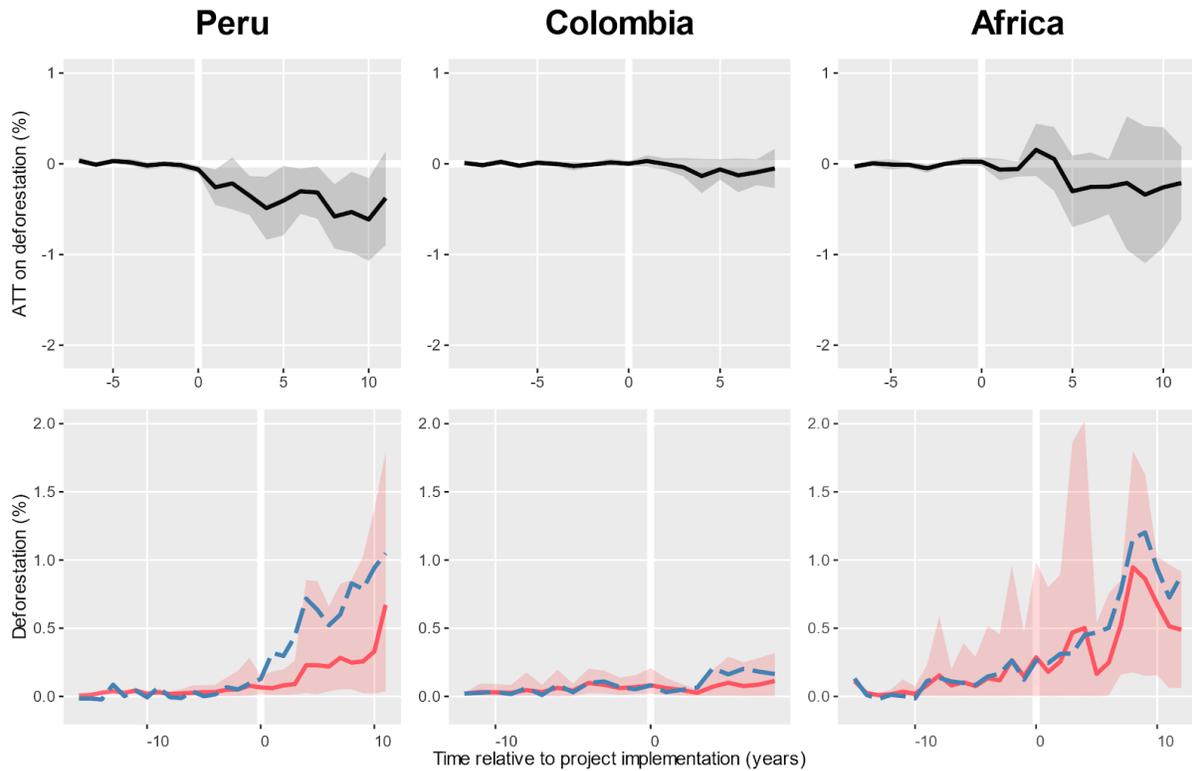

Figure S6. Estimated average impacts of REDD+ projects from Peru, Colombia, and Africa on annual deforestation, using the generalized synthetic control (GSC) method and selected controls from the ***genetic matching***. Upper panels display the average treatment effect on the treated (ATT) project areas. Lower panels display projects' (solid red line) and counterfactuals' (dashed blue line) deforestation averages. Shaded red areas represent bootstrapped 95% confidence intervals around the projects' deforestation average.

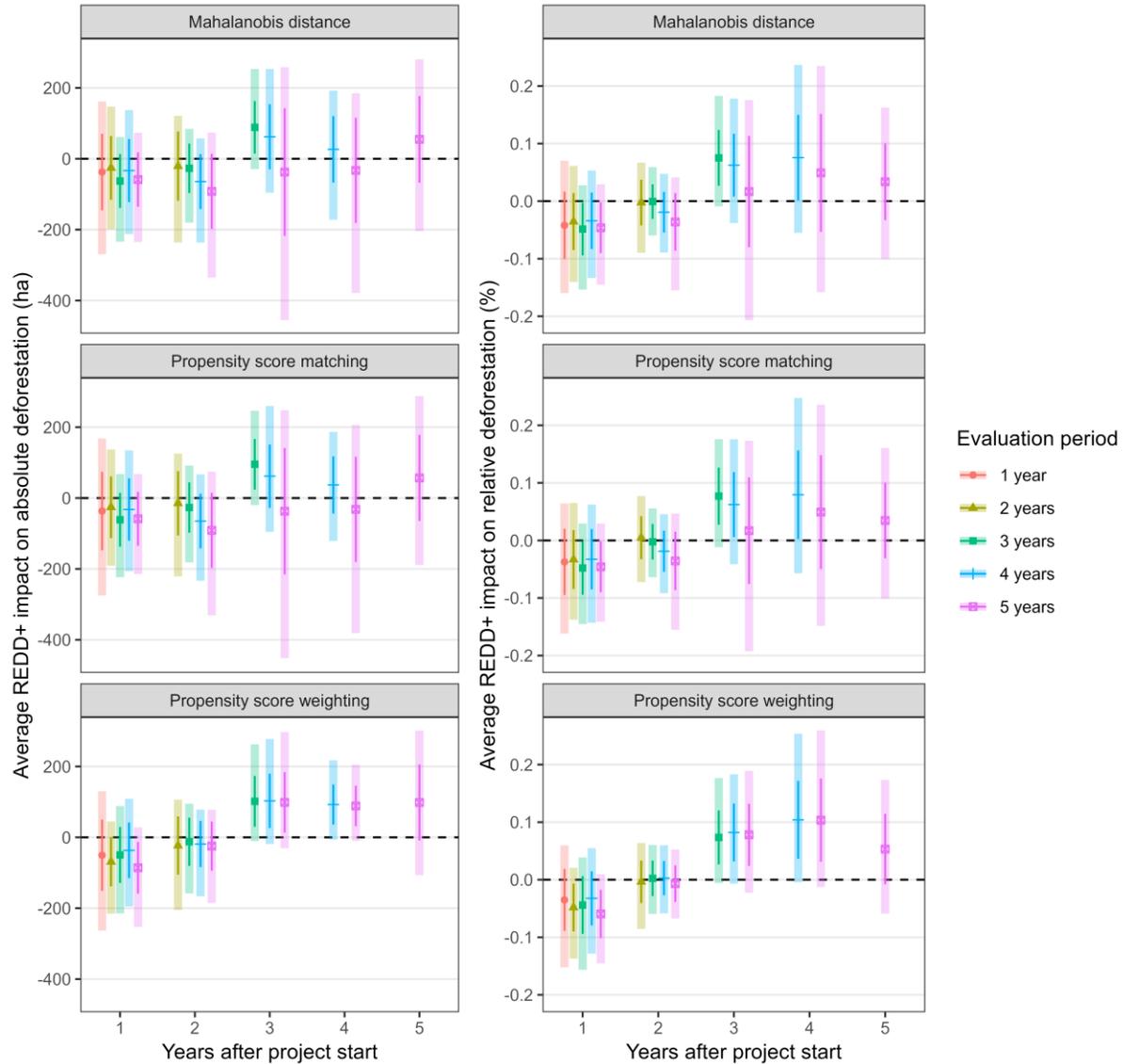

Figure S7. Estimated impacts of the REDD+ projects on the annual absolute and relative deforestation per evaluation period based on Mahalanobis distance, propensity score matching, and propensity score weighting (see Imai et al.[1] for details). Dark colored lines represent standard error bars. Light colored bars represent bootstrapped 95% confidence intervals. Impact estimates for the same year after project start vary across the evaluation periods due to changes in sample sizes (cf., West et al.[2]).

---

[1] Imai, K., Kim, I.S., Wang, E., 2018. Matching Methods for Causal Inference With Time-Series Cross-sectional Data. Harvard University, Massachusetts Institute of Technology, Cambridge.

[2] West, T.A.P., Caviglia-Harris, J.L., Martins, F.R.V., Silva, D.E., Börner, J., 2022. Potential conservation gains from improved protected area management in the Brazilian Amazon. Biological Conservation 269, 109526.

Table S10. Differences in the expected volume of carbon offsets generated by the voluntary REDD+ projects: observed cumulative deforestation in the project areas versus project baselines (ex-ante) versus synthetic control deforestation (ex-post) through 2020.

| Country | Project ID | Observed deforestation in the project area (ha) | Project baseline deforestation* (ex-ante; ha) | Observed deforestation in the synthetic control area (ex-post; ha) | Expected carbon offsets based on projects' ex-ante baseline† (Mg $CO_2$) | Proportional carbon offsets based on the synthetic control deforestation (Mg $CO_2$) | Evidence of significant reductions in deforestation‡ | Avoided deforestation based on the synthetic control (ha) | Carbon offsets based on the synthetic control (Mg $CO_2$) |
|---|---|---|---|---|---|---|---|---|---|
| Peru | 1882 | 403 | 1268 | 702 | 356,960 | 197,623 | Yes | 299 | 84,057 |
| | 2278 | 878 | 13299 | 873 | 5,891,253 | 386,726 | No | 0 | 0 |
| | 1067 | 4070 | 13581 | 8797 | 4,817,471 | 3,120,484 | Yes | 4727 | 1,676,658 |
| | 958 | 3391 | 2924 | 4653 | 630,937 | 1,004,018 | Yes | 1262 | 272,313 |
| | 944 | 5454 | 23685 | 8783 | 5,151,165 | 1,910,183 | Yes | 3329 | 724,012 |
| | 844 | 699 | 125541 | 410 | 12,475,134 | 40,742 | No | 0 | 0 |
| Colombia | 1400 | 334 | 4450 | 264 | 1,657,098 | 98,299 | No | 0 | 0 |
| | 1566 | 22,473 | 103,908 | 21,228 | 31,325,923 | 6,399,752 | No | 0 | 0 |
| | 1396 | 693 | 8076 | 6219 | 1,489,786 | 1,147,190 | Yes | 5526 | 1,019,305 |
| | 1395 | 1129 | 16,835 | 1253 | 2,791,723 | 207,787 | No | 124 | 20,616 |
| | 1392 | 105 | 10,722 | 118 | 1,455,141 | 16,014 | No | 0 | 0 |

Table S10 (continued). Differences in the expected volume of carbon offsets generated by the voluntary REDD+ projects: observed cumulative deforestation in the project areas versus project baselines (ex-ante) versus synthetic control deforestation (ex-post) through 2020.

| Country | Project ID | Observed deforestation in the project area (ha) | Project baseline deforestation* (ex-ante; ha) | Observed deforestation in the synthetic control area (ex-post; ha) | Expected carbon offsets based on projects' ex-ante baseline† (Mg CO$_2$) | Proportional carbon offsets based on the synthetic control deforestation (Mg CO$_2$) | Evidence of significant reductions in deforestation‡ | Avoided deforestation based on the synthetic control (ha) | Carbon offsets based on the synthetic control (Mg CO$_2$) |
|---|---|---|---|---|---|---|---|---|---|
| Cambodia | 904 | 12,950 | 21,252 | 10,632 | 1,626,420 | 813,669 | No | 0 | 0 |
| | 1650 | 11,499 | 30,446 | 15,609 | 12,432,277 | 6,373,757 | Yes | 4110 | 1,678,272 |
| DRC | 1359 | 16,385 | 11,949 | 8,204 | 4,735,361 | 3,251,226 | No | 0 | 0 |
| Tanzania | 1325 | 11,318 | 10,578 | 8,077 | 359,834 | 274,757 | No | 0 | 0 |
| | 1900 | 16 | 11,407 | 14 | 348,019 | 427 | No | 0 | 0 |
| | 1897 | 14161 | 35,472 | 13,577 | 1,406,892 | 538,492 | No | 0 | 0 |
| TOTAL | | | | | 88,951,394 | 25,781,146 | | | 5,475,233 |

\* Reported by official project documents.
† Reported by official project documents exclusively for the projects' avoided deforestation and forest degradation.
‡ Based on the results from the synthetic control analyses and placebo tests.
Note: Projects 856 and 1390 from Colombia, 934 from DRC, 1748 from Cambodia, and the Zambian projects and were excluded due to insufficient public information about ex-ante baseline deforestation rates. Projects with baselines ending before 2020 (e.g., 2018) were compared to their respective observed and synthetic control cumulative deforestations for the last available baseline year.

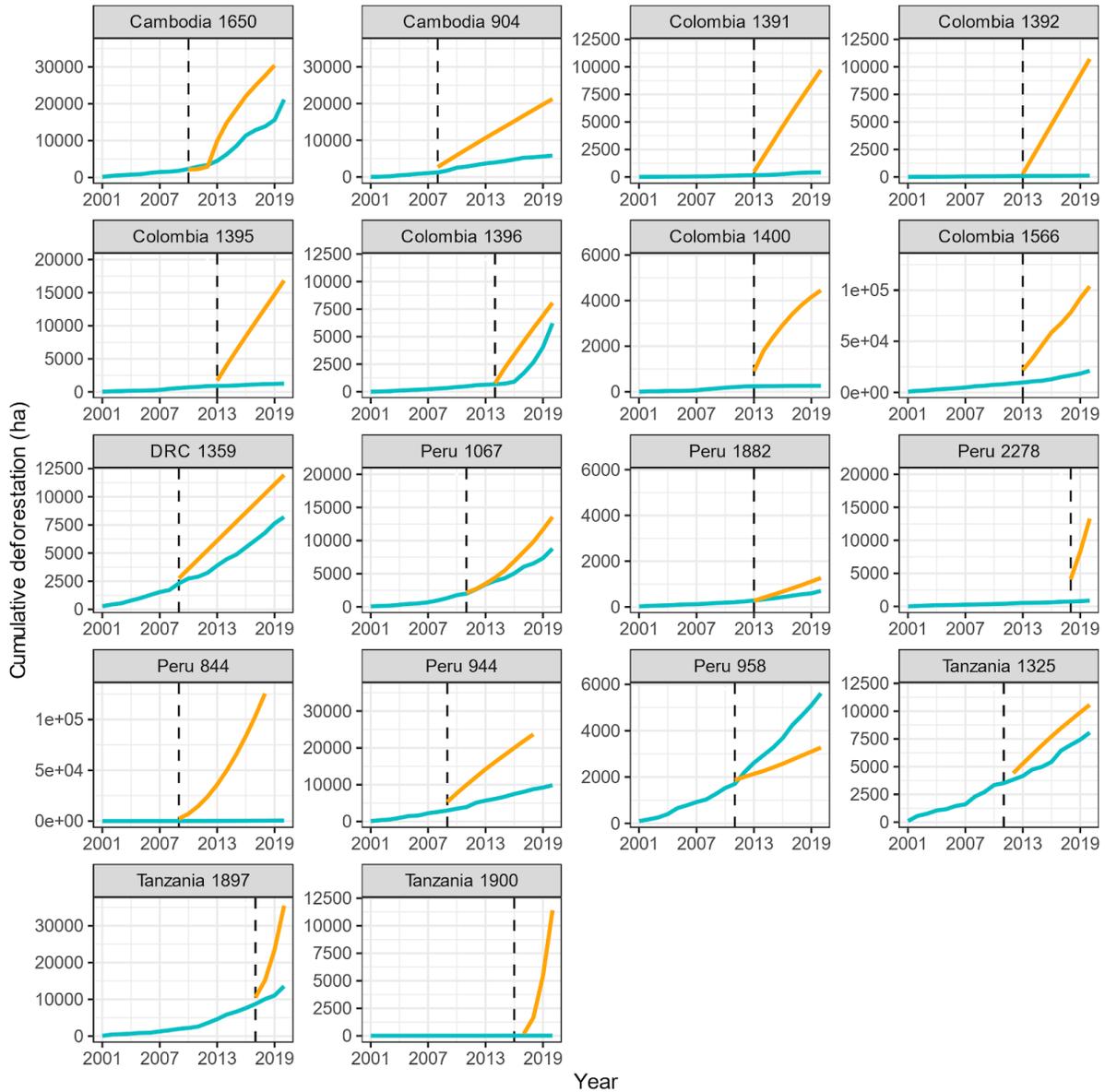

Figure S8. Cumulative deforestation from the baseline scenarios adopted by the REDD+ projects (orange) versus observed cumulative deforestation in the synthetic controls (blue). Dashed black lines indicate the project implementation year (note: scales differ).

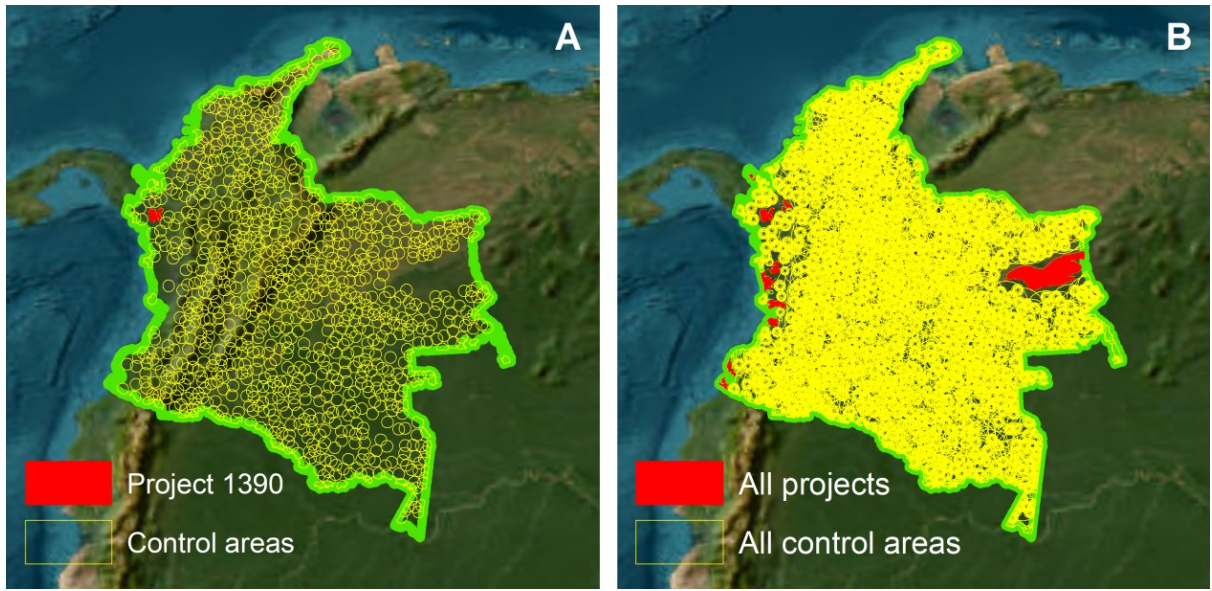

Figure S9. Example of the 1000 circular spatial polygons created randomly as potential control areas for Project #1390 in Colombia (A) and all combined polygons for the Colombian projects (B). Circular spatial polygons are later intersected with all covariate maps described in Table S8.

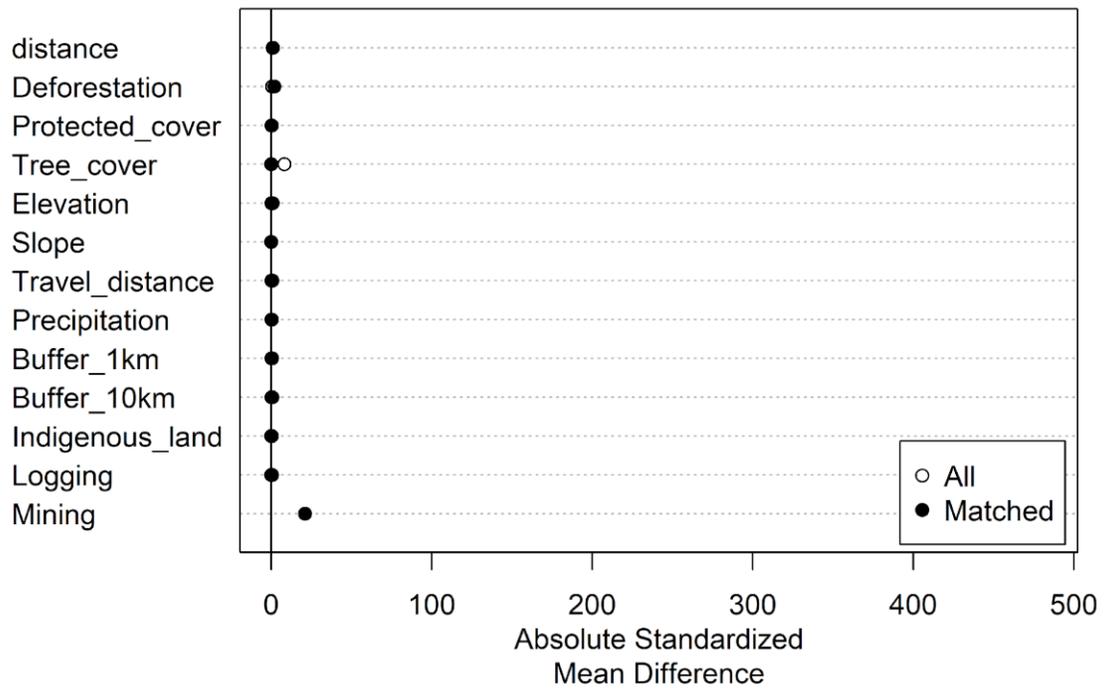

Figure S10. Covariate balance for the ***Peruvian projects*** before and after genetic matching. Absolute standardized mean differences within the vertical-line interval generally indicate well-balanced sets.

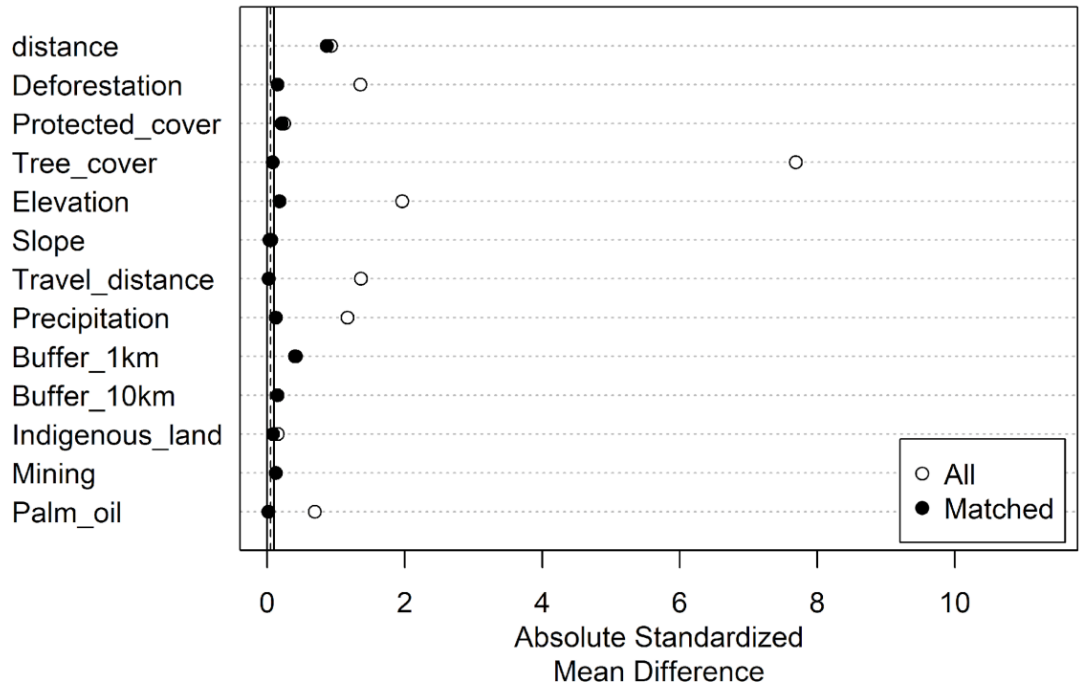

Figure S11. Covariate balance for the ***Colombian projects*** before and after genetic matching. Absolute standardized mean differences within the vertical-line interval generally indicate well-balanced sets.

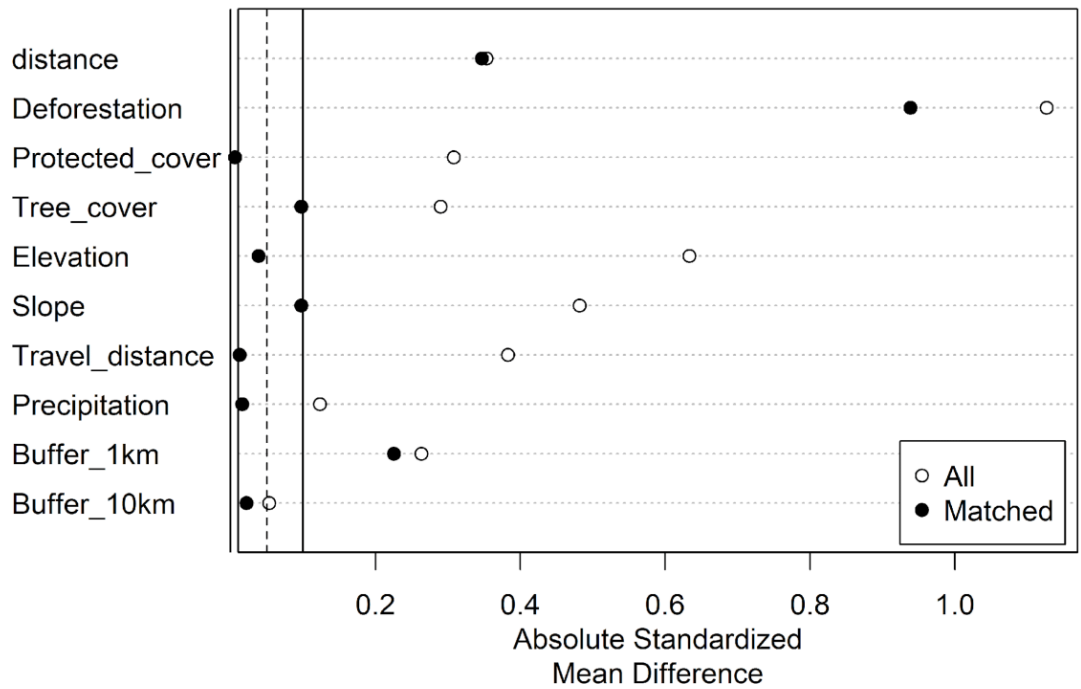

Figure S12. Covariate balance for the *African projects* before and after genetic matching. Absolute standardized mean differences within the vertical-line interval generally indicate well-balanced sets.

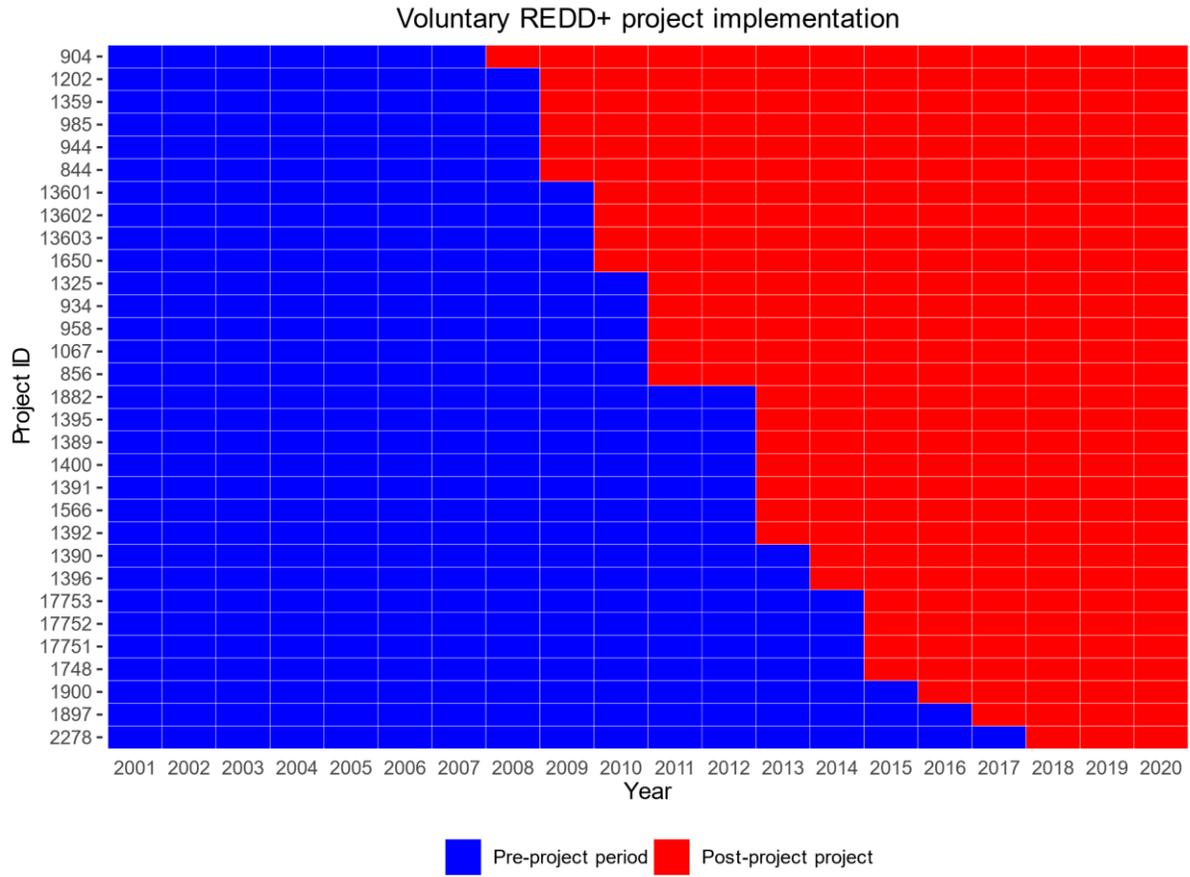

Figure S13. Treatment history: VCS-certified REDD+ project implementation.

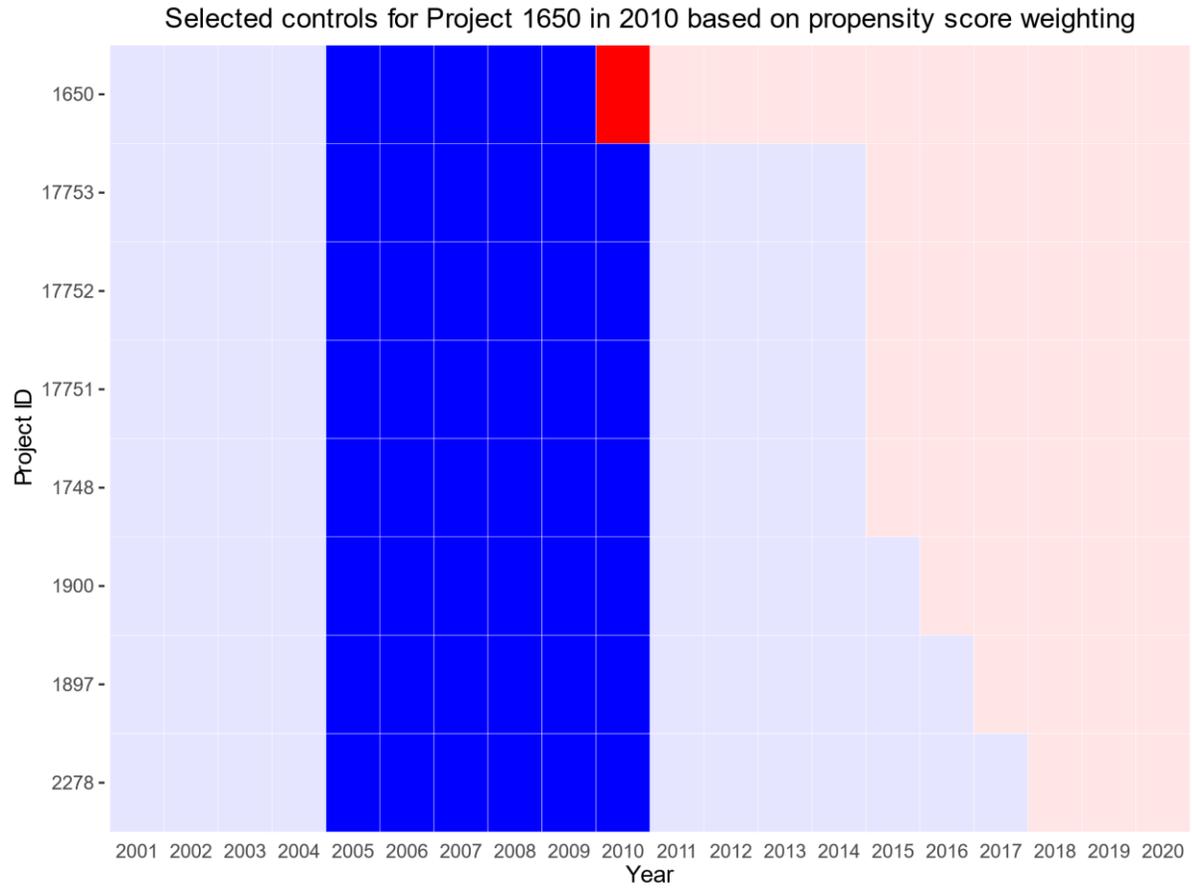

Figure S14. Example of selected controls ("to-be" REDD+ sites; blue) for the evaluation of operational Project 1650 in 2010 (red). Highlighted 2005–2009 period (blue) is the interval used for the identification of the control areas via matching.

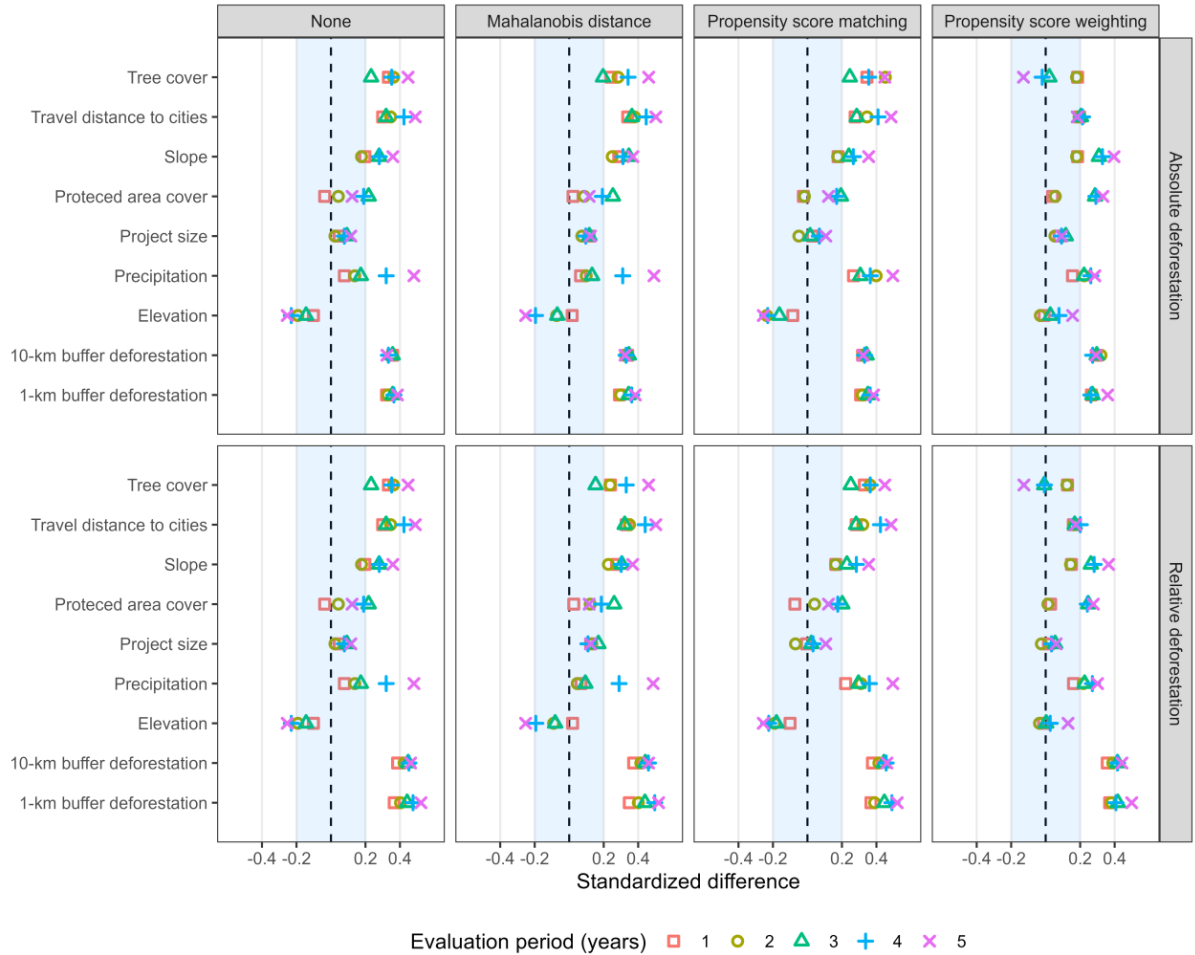

Figure S15. Covariate balance from propensity score matching, Mahalanobis distance, and propensity score weighting based on absolute (ha year$^{-1}$) and relative deforestation (% year$^{-1}$) for different evaluation periods (years). Time-varying covariates are represented by averages over the 5-year interval prior to project implementation used for the matching analyses (Fig. S17). "None" represents the original unmatched sample. Standardized differences within the shaded intervals generally indicate well-balanced sets.

Table S11. Data description.

| Variable | Description | Geographical coverage | Source |
|---|---|---|---|
| Deforestation and average tree cover in 2000 | 2000–2020 forest cover and change, time-variant | Global | Global Forest Change dataset (Hansen et al., 2013), available in Google Earth Engine |
| Travel time to urban centers | Friction surface of land-based travel time (days); time-invariant | Global | Global Friction Surface 2019 (Weiss et al., 2018), available in Google Earth Engine |
| Elevation and slope | Average elevation and slope in project and control areas (m); time-invariant | Global | Global Multi-resolution Terrain Elevation Data (GMTED2010), available in Google Earth Engine |
| Protected area cover | Location of protected areas; time-invariant | Peru | National Service of Natural Areas Protected by the State (SERNANP; Spanish acronym) |
| | | Colombia | Colombian Environmental Information System (SIAC; Spanish acronym) |
| | | Cambodia | Open Development Cambodia (https://opendevelopmentcambodia.net/) |
| | | DRC, Tanzania, Zambia | Protected Planet (https://www.protectedplanet.net/) |
| Indigenous land cover | Location of Indigenous lands; time-invariant | Peru | Body for the Formalization of Informal Property (COFOPRI; Spanish acronym) |
| | | Colombia | National Land Agency (ANT; Spanish acronym) |
| | | DRC | Map for Environment (https://mapforenvironment.org/) |
| Mining concession cover | Location of mining concessions; time-invariant | Peru, Colombia, DRC | Global Forest Watch (https://www.globalforestwatch.org/) |
| | | Cambodia | Open Development Cambodia (https://opendevelopmentcambodia.net/) |
| | | Tanzania | Map for Environment (https://mapforenvironment.org/) |
| Logging concession cover | Location of logging concessions; time-invariant | Peru | Map for Environment (https://mapforenvironment.org/) |
| | | DRC | Global Forest Watch (https://www.globalforestwatch.org/) |
| Palm oil concession cover | Location of palm oil concessions; time-invariant | Colombia | Map for Environment (https://mapforenvironment.org/) |
| | | Indonesia | Global Forest Watch (https://www.globalforestwatch.org/) |
| General concession cover | Location of multiple land concessions; time-invariant | Cambodia | Map for Environment (https://mapforenvironment.org/) |
| Precipitation | Annual cumulative precipitation (mm); time-variant | Global | Monthly Global Precipitation Measurement (GPM) v.6, available in Google Earth Engine |
| REDD+ jurisdictional programs and other conservation interventions | Central Africa Regional Program for the Environment (CARPE) in RDC, DRC's jurisdictional Maï Ndombe Emission Reduction Program, and DRC's Forest Investment Program; time-invariant | DRC | Map for Environment (https://mapforenvironment.org/) |
| Other REDD+ interventions | Other REDD+ intervention areas used for exclusion of control areas; time-variant | DRC | Map for Environment (https://mapforenvironment.org/) |
| Soil fertility | 3-class ordinal map of soil fertility; time-invariant | Cambodia | Map for Environment (https://mapforenvironment.org/) |

**Appendix A. Optimized v-weights from the synthetic control (SC) analyses and covariate balance between the voluntary REDD+ project sites and SCs from 2001 to project implementation year (note: missing covariates indicate no covariate variation between the project and control areas; see Table S10).**

Table A1. Covariate balance for Project 844 from Peru.

| Covariate | V-weights | Project area | Synthetic control* | Control set average |
|---|---|---|---|---|
| Average tree cover (%) | 5.00E-09 | 0.99 | 0.98 | 0.813 |
| Indigenous land cover (%) | 5.00E-09 | 0 | 0.009 | 0.168 |
| Protected area cover (%) | 5.00E-09 | 0 | 0.43 | 0.184 |
| Mining concession cover (%) | 0.499943 | 0 | 0 | 0.027 |
| Slope (degree) | 0.499943 | 2.271 | 2.271 | 11.609 |
| Elevation (m) | 5.00E-09 | 331.537 | 232.917 | 1142.592 |
| Timber concession cover (%) | 5.00E-09 | 1 | 0.518 | 0.126 |
| Travel distance to urban centers (days) | 8.07E-06 | 0.038 | 0.038 | 0.05 |
| Average annual deforestation (ha) | 5.00E-09 | 6.785 | 6.38 | 87.424 |
| Average cumulative deforestation (ha) | 0.000106 | 25.668 | 25.987 | 369.073 |
| Annual precipitation (mm) | 1.17E-07 | 1874 | 2305.455 | 1932.274 |
| Average annual buffer deforestation (ha) | 5.00E-09 | 139.75 | 129.97 | 138.387 |
| Average cumulative buffer deforestation (ha) | 7.65E-08 | 445.125 | 502.823 | 584.46 |

*Based on four control areas.

Table A2. Covariate balance for Project 944 from Peru.

| Covariate | V-weights | Project area | Synthetic control* | Control set average |
|---|---|---|---|---|
| Average tree cover (%) | 1.00E-08 | 0.89 | 0.96 | 0.903 |
| Indigenous land cover (%) | 1.00E-08 | 0.011 | 0.341 | 0.14 |
| Protected area cover (%) | 1.00E-08 | 0.98 | 0.004 | 0.075 |
| Mining concession cover (%) | 1.00E-08 | 0 | 0.266 | 0.073 |
| Slope (degree) | 1.00E-08 | 17.249 | 10.755 | 8.95 |
| Elevation (m) | 1.00E-08 | 1751.28 | 1031.965 | 652.62 |
| Timber concession cover (%) | 1.00E-08 | 0 | 0.019 | 0.096 |
| Travel distance to urban centers (days) | 1.00E-08 | 0.076 | 0.056 | 0.049 |
| Average annual deforestation (ha) | 1 | 304.811 | 326.58 | 1133.672 |
| Average cumulative deforestation (ha) | 1.00E-08 | 1069.94 | 1252.255 | 4711.364 |
| Annual precipitation (mm) | 1.00E-08 | 1208.125 | 1665.75 | 1705.13 |
| Average annual buffer deforestation (ha) | 1.00E-08 | 1444.625 | 1157.125 | 1382.606 |
| Average cumulative buffer deforestation (ha) | 1.00E-08 | 5440.25 | 4319 | 5662.537 |

*Based on one control area.

Table A3. Covariate balance for Project 958 from Peru.

| Covariate | V-weights | Project area | Synthetic control* | Control set average |
|---|---|---|---|---|
| Average tree cover (%) | 9.30E-09 | 0.93 | 0.734 | 0.837 |
| Indigenous land cover (%) | 0.003971 | 0 | 0.088 | 0.196 |
| Protected area cover (%) | 9.30E-09 | 0 | 0.034 | 0.138 |
| Slope (degree) | 9.30E-09 | 23.243 | 20.099 | 10.722 |
| Elevation (m) | 9.30E-09 | 1831.213 | 2256.768 | 1061 |
| Mining concession cover (%) | 0.060823 | 0 | 0.021 | 0.063 |
| Timber concession cover (%) | 0.000219 | 1 | 0.202 | 0.088 |
| Travel distance to urban centers (days) | 0.003126 | 0.087 | 0.075 | 0.051 |
| Average annual deforestation (ha) | 9.30E-09 | 158.09 | 152.822 | 550.58 |
| Average cumulative deforestation (ha) | 0.929849 | 712.794 | 711.536 | 2803.392 |
| Annual precipitation (mm) | 9.30E-09 | 1139.8 | 1397.511 | 1830.01 |
| Average annual buffer deforestation (ha) | 9.30E-09 | 482.1 | 479.555 | 462.755 |
| Average cumulative buffer deforestation (ha) | 0.002012 | 2305 | 2425.142 | 2324.838 |

*Based on four control areas.

Table A4. Covariate balance for Project 1067 from Peru.

| Covariate | V-weights | Project area | Synthetic control* | Control set average |
|---|---|---|---|---|
| Average tree cover (%) | 8.21E-09 | 0.98 | 0.985 | 0.844 |
| Indigenous land cover (%) | 8.21E-09 | 0.002 | 0.019 | 0.19 |
| Protected area cover (%) | 8.21E-09 | 0.939 | 0.014 | 0.141 |
| Slope (degree) | 8.21E-09 | 0.929 | 1.604 | 11.508 |
| Elevation (m) | 8.21E-09 | 234.969 | 186.731 | 1120.772 |
| Mining concession cover (%) | 8.21E-09 | 0.001 | 0 | 0.054 |
| Timber concession cover (%) | 8.21E-09 | 0.124 | 0.731 | 0.108 |
| Travel distance to urban centers (days) | 7.71E-06 | 0.038 | 0.039 | 0.052 |
| Average annual deforestation (ha) | 0.178866 | 154.182 | 176.482 | 1374.262 |
| Average cumulative deforestation (ha) | 0.821126 | 658.882 | 642.699 | 6784.841 |
| Annual precipitation (mm) | 8.21E-09 | 2825.3 | 3309.608 | 1841.981 |
| Average annual buffer deforestation (ha) | 8.21E-09 | 1172.9 | 705.164 | 972.285 |
| Average cumulative buffer deforestation (ha) | 8.21E-09 | 5819.4 | 3352.26 | 4756.742 |

*Based on two control areas.

Table A5. Covariate balance for Project 1882 from Peru.

| Covariate | V-weights | Project area | Synthetic control* | Control set average |
|---|---|---|---|---|
| Average tree cover (%) | 1.63E-06 | 0.85 | 0.868 | 0.795 |
| Indigenous land cover (%) | 5.85E-09 | 0 | 0.003 | 0.179 |
| Protected area cover (%) | 0.413688 | 0 | 0 | 0.144 |
| Slope (degree) | 5.85E-09 | 26.758 | 5.291 | 10.457 |
| Elevation (m) | 1.97E-08 | 2855.566 | 507.714 | 1073.627 |
| Mining concession cover (%) | 9.29E-06 | 0 | 0.004 | 0.057 |
| Timber concession cover (%) | 5.85E-09 | 1 | 0.301 | 0.115 |
| Travel distance to urban centers (days) | 2.70E-08 | 0.091 | 0.037 | 0.046 |
| Average annual deforestation (ha) | 0.585399 | 20.024 | 20.024 | 46.98 |
| Average cumulative deforestation (ha) | 0.000901 | 123.477 | 123.016 | 252.864 |
| Annual precipitation (mm) | 1.29E-08 | 1319.333 | 2858.851 | 2034.109 |
| Average annual buffer deforestation (ha) | 5.85E-09 | 115.333 | 117.029 | 113.614 |
| Average cumulative buffer deforestation (ha) | 5.85E-09 | 677 | 713.303 | 625.602 |

*Based on six control areas.

Table A6. Covariate balance for Project 2278 from Peru.

| Covariate | V-weights | Project area | Synthetic control* | Control set average |
|---|---|---|---|---|
| Average tree cover (%) | 1.00E-08 | 0.99 | 0.985 | 0.862 |
| Indigenous land cover (%) | 1.00E-08 | 0 | 0.138 | 0.218 |
| Protected area cover (%) | 1.00E-08 | 0 | 0.001 | 0.133 |
| Slope (degree) | 1.00E-08 | 2.153 | 1.766 | 10.155 |
| Elevation (m) | 1.00E-08 | 361.254 | 277.825 | 949.285 |
| Mining concession cover (%) | 0.999878 | 0 | 0.029 | 0.06 |
| Timber concession cover (%) | 1.00E-08 | 1 | 0.792 | 0.122 |
| Travel distance to urban centers (days) | 1.00E-08 | 0.039 | 0.038 | 0.051 |
| Average annual deforestation (ha) | 1.00E-08 | 38.871 | 40.513 | 538.241 |
| Average cumulative deforestation (ha) | 0.000122 | 346.858 | 346.853 | 3803.924 |
| Annual precipitation (mm) | 1.00E-08 | 1980.765 | 3116.82 | 1891.379 |
| Average annual buffer deforestation (ha) | 1.00E-08 | 749.941 | 248.545 | 580.549 |
| Average cumulative buffer deforestation (ha) | 1.00E-08 | 5171.529 | 1376.397 | 4031.998 |

*Based on three control areas.

Table A7. Covariate balance for Project 1360-1 from Peru.

| Covariate | V-weights | Project area | Synthetic control* | Control set average |
|---|---|---|---|---|
| Average tree cover (%) | 1.59E-07 | 0.99 | 0.96 | 0.899 |
| Indigenous land cover (%) | 1.00E-08 | 0.998 | 0.917 | 0.156 |
| Protected area cover (%) | 1 | 0 | 0 | 0.067 |
| Mining concession cover (%) | 1.00E-08 | 0 | 0.205 | 0.038 |
| Slope (degree) | 1.00E-08 | 2.47 | 12.623 | 6.281 |
| Elevation (m) | 1.00E-08 | 277.081 | 1131.4 | 488.26 |
| Timber concession cover (%) | 1.00E-08 | 0 | 0 | 0.082 |
| Travel distance to urban centers (days) | 1.00E-08 | 0.038 | 0.06 | 0.043 |
| Average annual deforestation (ha) | 1.00E-08 | 28.031 | 85.193 | 806.992 |
| Average cumulative deforestation (ha) | 1.00E-08 | 104.729 | 310.309 | 3484.132 |
| Annual precipitation (mm) | 1.00E-08 | 3398 | 1508.889 | 1957.044 |
| Average annual buffer deforestation (ha) | 1.00E-08 | 2764.444 | 1071.444 | 1320.204 |
| Average cumulative buffer deforestation (ha) | 1.00E-08 | 12250.56 | 3886 | 5731.282 |

*Based on one control areas.

Table A8. Covariate balance for Project 1360-2 from Peru.

| Covariate | V-weights | Project area | Synthetic control* | Control set average |
|---|---|---|---|---|
| Average tree cover (%) | 9.99E-09 | 0.99 | 1 | 0.916 |
| Indigenous land cover (%) | 9.99E-09 | 0.998 | 0.349 | 0.226 |
| Protected area cover (%) | 4.07E-06 | 0 | 0.001 | 0.114 |
| Mining concession cover (%) | 0.998817 | 0 | 0 | 0.031 |
| Slope (degree) | 9.99E-09 | 0.634 | 1.723 | 8.815 |
| Elevation (m) | 2.42E-07 | 169 | 197.704 | 738.592 |
| Timber concession cover (%) | 9.99E-09 | 0.001 | 0 | 0.133 |
| Travel distance to urban centers (days) | 9.99E-09 | 0.037 | 0.038 | 0.051 |
| Average annual deforestation (ha) | 7.01E-05 | 1.859 | 1.58 | 66.026 |
| Average cumulative deforestation (ha) | 0.001108 | 5.266 | 5.405 | 300.644 |
| Annual precipitation (mm) | 9.99E-09 | 1625.778 | 2156.448 | 1947.459 |
| Average annual buffer deforestation (ha) | 9.99E-09 | 292 | 213.611 | 282.821 |
| Average cumulative buffer deforestation (ha) | 9.99E-09 | 1307.667 | 932.363 | 1270.729 |

*Based on two control areas.

Table A9. Covariate balance for Project 1360-3 from Peru.

| Covariate | V-weights | Project area | Synthetic control* | Control set average |
|---|---|---|---|---|
| Average tree cover (%) | 1.00E-08 | 0.99 | 0.977 | 0.906 |
| Indigenous land cover (%) | 1.00E-08 | 0.989 | 0.203 | 0.171 |
| Protected area cover (%) | 0.99986 | 0 | 0 | 0.024 |
| Mining concession cover (%) | 1.00E-08 | 0 | 0.821 | 0.117 |
| Slope (degree) | 1.00E-08 | 0.726 | 4.336 | 7.854 |
| Elevation (m) | 1.00E-08 | 183.504 | 412.574 | 578.512 |
| Timber concession cover (%) | 1.00E-08 | 0 | 0.431 | 0.134 |
| Travel distance to urban centers (days) | 5.65E-08 | 0.037 | 0.042 | 0.046 |
| Average annual deforestation (ha) | 0.00014 | 7.757 | 9.204 | 89.47 |
| Average cumulative deforestation (ha) | 1.00E-08 | 25.114 | 44.374 | 410.975 |
| Annual precipitation (mm) | 1.00E-08 | 1641.556 | 4526.388 | 2305.322 |
| Average annual buffer deforestation (ha) | 1.00E-08 | 491.333 | 458.916 | 477.606 |
| Average cumulative buffer deforestation (ha) | 1.00E-08 | 1545.556 | 2017.51 | 2184.875 |

*Based on two control areas.

Table A10. Covariate balance for Project 856 from Colombia.

| Covariate | V-weights | Project area | Synthetic control* | Control set average |
|---|---|---|---|---|
| Average tree cover (%) | 6.33E-07 | 0.95 | 0.919 | 0.611 |
| Indigenous land cover (%) | 8.58E-09 | 0.003 | 0 | 0.048 |
| Protected area cover (%) | 4.30E-08 | 0.961 | 0.744 | 0.1 |
| Slope (degree) | 8.58E-09 | 14.87 | 9.171 | 5.132 |
| Elevation (m) | 8.58E-09 | 443.552 | 655.756 | 440.963 |
| Travel distance to urban centers (days) | 8.58E-09 | 0.061 | 0.055 | 0.024 |
| Average annual deforestation (ha) | 0.857906 | 13.448 | 13.448 | 50.482 |
| Average cumulative deforestation (ha) | 0.000324 | 58.197 | 58.602 | 294.724 |
| Annual precipitation (mm) | 0.141769 | 2837.455 | 2837.454 | 2669.489 |
| Average annual buffer deforestation (ha) | 8.58E-09 | 332 | 301.513 | 318.576 |
| Average cumulative buffer deforestation (ha) | 1.05E-07 | 1819.091 | 1641.475 | 1881.325 |

*Based on six control areas.

Table A11. Covariate balance for Project 1389 from Colombia.

| Covariate | V-weights | Project area | Synthetic control* | Control set average |
|---|---|---|---|---|
| Average tree cover (%) | 5.00E-09 | 0.95 | 0.99 | 0.71 |
| Indigenous land cover (%) | 0.5 | 0 | 0 | 0.015 |
| Protected area cover (%) | 5.00E-09 | 0.104 | 0 | 0.127 |
| Slope (degree) | 5.00E-09 | 1.567 | 1.736 | 2.661 |
| Elevation (m) | 5.00E-09 | 19.109 | 250.352 | 253.684 |
| Palm oil concession cover (%) | 5.00E-09 | 0.114 | 0 | 0.042 |
| Mining concession cover (%) | 0.5 | 0 | 0 | 0.004 |
| Travel distance to urban centers (days) | 5.00E-09 | 0.037 | 0.038 | 0.023 |
| Average annual deforestation (ha) | 5.00E-09 | 72.391 | 157.89 | 630.371 |
| Average cumulative deforestation (ha) | 3.75E-07 | 509.116 | 835.664 | 4138.333 |
| Annual precipitation (mm) | 5.00E-09 | 3096.25 | 2862.75 | 2983.195 |
| Average annual buffer deforestation (ha) | 5.00E-09 | 1254.25 | 986.917 | 1231.789 |
| Average cumulative buffer deforestation (ha) | 5.00E-09 | 8147.583 | 4981.667 | 7995.052 |

*Based on one control area.

Table A12. Covariate balance for Project 1390 from Colombia.

| Covariate | V-weights | Project area | Synthetic control* | Control set average |
|---|---|---|---|---|
| Average tree cover (%) | 4.94E-09 | 0.94 | 0.967 | 0.576 |
| Indigenous land cover (%) | 4.94E-09 | 0.002 | 0.001 | 0.106 |
| Protected area cover (%) | 4.94E-09 | 0 | 0.626 | 0.133 |
| Slope (degree) | 2.34E-06 | 0.936 | 1.834 | 8.9 |
| Elevation (m) | 1.14E-07 | 23.101 | 255.41 | 1009.857 |
| Mining concession cover (%) | 0.494015 | 0 | 0 | 0.002 |
| Travel distance to urban centers (days) | 0.494015 | 0.038 | 0.037 | 0.032 |
| Average annual deforestation (ha) | 0.000999 | 28.59 | 28.374 | 178.222 |
| Average cumulative deforestation (ha) | 0.010968 | 161.489 | 161.673 | 1284.22 |
| Annual precipitation (mm) | 4.94E-09 | 4587 | 2662.707 | 2352.755 |
| Average annual buffer deforestation (ha) | 4.94E-09 | 235.231 | 196.72 | 233.208 |
| Average cumulative buffer deforestation (ha) | 4.94E-09 | 1595.692 | 1247.988 | 1688.764 |

*Based on four control areas.

Table A13. Covariate balance for Project 1391 from Colombia.

| Covariate | V-weights | Project area | Synthetic control* | Control set average |
|---|---|---|---|---|
| Average tree cover (%) | 1.73E-05 | 0.98 | 0.977 | 0.623 |
| Indigenous land cover (%) | 1.00E-08 | 0.007 | 0.619 | 0.293 |
| Protected area cover (%) | 1.00E-08 | 0.513 | 0 | 0.16 |
| Slope (degree) | 1.00E-08 | 4.314 | 1.661 | 6.757 |
| Elevation (m) | 1.79E-06 | 56.516 | 120.435 | 1020.077 |
| Palm oil concession cover (%) | 1.00E-08 | 0.166 | 0 | 0.005 |
| Mining concession cover (%) | 0.999972 | 0 | 0 | 0.013 |
| Travel distance to urban centers (days) | 1.00E-08 | 0.039 | 0.038 | 0.036 |
| Average annual deforestation (ha) | 2.09E-06 | 14.192 | 12.819 | 21.904 |
| Average cumulative deforestation (ha) | 6.72E-06 | 55.714 | 60.846 | 141.402 |
| Annual precipitation (mm) | 1.00E-08 | 6758.25 | 4982.011 | 2918.984 |
| Average annual buffer deforestation (ha) | 1.00E-08 | 52.5 | 47.307 | 52.257 |
| Average cumulative buffer deforestation (ha) | 1.00E-08 | 237.833 | 288.669 | 341.149 |

*Based on two control areas.

Table A14. Covariate balance for Project 1392 from Colombia.

| Covariate | V-weights | Project area | Synthetic control* | Control set average |
|---|---|---|---|---|
| Average tree cover (%) | 1.91E-07 | 0.98 | 0.588 | 0.687 |
| Indigenous land cover (%) | 1.30E-06 | 0 | 0.095 | 0.47 |
| Protected area cover (%) | 1.77E-07 | 0.014 | 0.363 | 0.185 |
| Slope (degree) | 0.998654 | 6.369 | 6.369 | 2.42 |
| Elevation (m) | 1.77E-07 | 65.338 | 816.682 | 256.4 |
| Travel distance to urban centers (days) | 2.12E-07 | 0.039 | 0.033 | 0.03 |
| Average annual deforestation (ha) | 0.000472 | 6.984 | 7.045 | 6.786 |
| Average cumulative deforestation (ha) | 0.000872 | 42.199 | 42.085 | 38.622 |
| Annual precipitation (mm) | 1.77E-07 | 5160.333 | 3701.342 | 3215.53 |
| Average annual buffer deforestation (ha) | 1.77E-07 | 13.75 | 13.663 | 13.712 |
| Average cumulative buffer deforestation (ha) | 1.77E-07 | 66.833 | 75.962 | 82.085 |

*Based on five control areas.

Table A15. Covariate balance for Project 1395 from Colombia.

| Covariate | V-weights | Project area | Synthetic control* | Control set average |
|---|---|---|---|---|
| Average tree cover (%) | 1.93E-06 | 0.98 | 0.965 | 0.893 |
| Indigenous land cover (%) | 1.00E-08 | 0 | 0.22 | 0.367 |
| Protected area cover (%) | 1.00E-08 | 0.279 | 0 | 0.113 |
| Slope (degree) | 1.00E-08 | 2.891 | 3.175 | 7.928 |
| Elevation (m) | 1.00E-08 | 53.587 | 100.28 | 888.254 |
| Palm oil concession cover (%) | 1.00E-08 | 0.377 | 0.076 | 0.02 |
| Mining concession cover (%) | 0.999938 | 0 | 0 | 0.013 |
| Travel distance to urban centers (days) | 1.00E-08 | 0.038 | 0.039 | 0.047 |
| Average annual deforestation (ha) | 5.72E-05 | 72.962 | 72.719 | 63.174 |
| Average cumulative deforestation (ha) | 2.67E-06 | 352.45 | 380.712 | 394.998 |
| Annual precipitation (mm) | 1.00E-08 | 6213.583 | 6167.6 | 3092.443 |
| Average annual buffer deforestation (ha) | 1.00E-08 | 123.667 | 138.386 | 125.305 |
| Average cumulative buffer deforestation (ha) | 3.48E-08 | 552.833 | 704.622 | 803.245 |

*Based on two control areas.

Table A16. Covariate balance for Project 1396 from Colombia.

| Covariate | V-weights | Project area | Synthetic control* | Control set average |
|---|---|---|---|---|
| Average tree cover (%) | 1.18E-08 | 0.98 | 0.761 | 0.487 |
| Indigenous land cover (%) | 6.69E-06 | 0.001 | 0 | 0.046 |
| Protected area cover (%) | 0.440264 | 0 | 0 | 0.097 |
| Slope (degree) | 5.24E-08 | 3.318 | 2.001 | 7.828 |
| Elevation (m) | 4.40E-09 | 48.842 | 270.013 | 676.986 |
| Mining concession cover (%) | 0.440264 | 0 | 0 | 0.003 |
| Palm oil concession cover (%) | 4.40E-09 | 0.018 | 0.192 | 0.184 |
| Travel distance to urban centers (days) | 4.40E-09 | 0.039 | 0.03 | 0.023 |
| Average annual deforestation (ha) | 4.40E-09 | 47.55 | 48.627 | 132.717 |
| Average cumulative deforestation (ha) | 0.119465 | 269.257 | 269.257 | 939.447 |
| Annual precipitation (mm) | 8.17E-09 | 7466.846 | 3114.119 | 2359.675 |
| Average annual buffer deforestation (ha) | 4.40E-09 | 297 | 297.722 | 296.405 |
| Average cumulative buffer deforestation (ha) | 4.40E-09 | 1755.308 | 2090.06 | 2126.976 |

*Based on three control areas.

Table A17. Covariate balance for Project 1400 from Colombia.

| Covariate | V-weights | Project area | Synthetic control* | Control set average |
|---|---|---|---|---|
| Average tree cover (%) | 7.80E-09 | 0.97 | 0.936 | 0.571 |
| Indigenous land cover (%) | 7.80E-09 | 0.041 | 0.026 | 0.145 |
| Protected area cover (%) | 7.80E-09 | 0.5 | 0.423 | 0.137 |
| Slope (degree) | 7.80E-09 | 2.812 | 8.312 | 8.417 |
| Elevation (m) | 7.80E-09 | 50.619 | 1241.078 | 967.214 |
| Mining concession cover (%) | 7.80E-09 | 0 | 0.056 | 0.006 |
| Palm oil concession cover (%) | 7.80E-09 | 0.18 | 0.043 | 0.044 |
| Travel distance to urban centers (days) | 7.80E-09 | 0.038 | 0.056 | 0.032 |
| Average annual deforestation (ha) | 0.219686 | 20.939 | 19.509 | 63.069 |
| Average cumulative deforestation (ha) | 0.780314 | 88.433 | 92.332 | 407.686 |
| Annual precipitation (mm) | 7.80E-09 | 7152.667 | 4465.09 | 2584.668 |
| Average annual buffer deforestation (ha) | 7.80E-09 | 131.5 | 133.086 | 131.12 |
| Average cumulative buffer deforestation (ha) | 7.80E-09 | 646.417 | 693.196 | 836.637 |

*Based on two control areas.

Table A18. Covariate balance for Project 1566 from Colombia.

| Covariate | V-weights | Project area | Synthetic control* | Control set average |
|---|---|---|---|---|
| Average tree cover (%) | 3.04E-09 | 0.85 | 0.807 | 0.512 |
| Indigenous land cover (%) | 3.04E-09 | 0.901 | 0.498 | 0.126 |
| Protected area cover (%) | 5.60E-06 | 0 | 0.009 | 0.113 |
| Slope (degree) | 0.304029 | 1.076 | 1.076 | 6.187 |
| Elevation (m) | 4.81E-05 | 128.519 | 169.262 | 813.145 |
| Mining concession cover (%) | 0.091708 | 0.002 | 0.002 | 0.001 |
| Travel distance to urban centers (days) | 3.04E-09 | 0.037 | 0.032 | 0.021 |
| Average annual deforestation (ha) | 5.20E-05 | 732.768 | 743.14 | 2296.36 |
| Average cumulative deforestation (ha) | 0.303551 | 4694.777 | 4694.767 | 14677.86 |
| Annual precipitation (mm) | 0.300605 | 2828 | 2828 | 2431.834 |
| Average annual buffer deforestation (ha) | 7.18E-09 | 653.75 | 585.906 | 662.36 |
| Average cumulative buffer deforestation (ha) | 3.04E-09 | 4758.5 | 3731.181 | 4264.713 |

*Based on six control areas.

Table A19. Covariate balance for Project 934 from the Democratic Republic of Congo.

| Covariate | V-weights | Project area | Synthetic control* | Control set average |
|---|---|---|---|---|
| Average tree cover (%) | 2.22E-08 | 0.89 | 0.589 | 0.639 |
| Indigenous land cover (%) | 0.835655 | 0 | 0 | 0.091 |
| Protected area cover (%) | 8.36E-09 | 0.787 | 2.623 | 2.361 |
| Slope (degree) | 8.36E-09 | 319.012 | 903.379 | 713.075 |
| Elevation (m) | 3.62E-08 | 0 | 0.642 | 0.441 |
| Mining concession cover (%) | 8.36E-09 | 0.697 | 0.016 | 0.25 |
| Logging concession cover (%) | 1.28E-08 | 1 | 0 | 0.051 |
| REDD+ jurisdiction or other conservation program(s) cover (%) | 3.21E-08 | 1 | 0.094 | 0.184 |
| Travel distance to urban centers (days) | 8.36E-09 | 0.038 | 0.022 | 0.024 |
| Average annual deforestation (ha) | 4.88E-06 | 797.509 | 764.554 | 360.083 |
| Average cumulative deforestation (ha) | 0.16434 | 3506.279 | 3506.282 | 1766.389 |
| Annual precipitation (mm) | 7.75E-08 | 1691.8 | 1380.623 | 1501.567 |
| Average annual buffer deforestation (ha) | 8.36E-09 | 294.2 | 329.185 | 291.748 |
| Average cumulative buffer deforestation (ha) | 8.36E-09 | 1345.8 | 1479.023 | 1429.909 |

*Based on four control areas.

Table A20. Covariate balance for Project 1359 from the Democratic Republic of Congo.

| Covariate | V-weights | Project area | Synthetic control* | Control set average |
|---|---|---|---|---|
| Average tree cover (%) | 1.00E-08 | 1 | 0.981 | 0.83 |
| Indigenous land cover (%) | 7.82E-05 | 0 | 0.255 | 0.013 |
| Protected area cover (%) | 1.00E-08 | 1.522 | 4.072 | 2.732 |
| Slope (degree) | 0.000153 | 432.489 | 631.843 | 556.565 |
| Elevation (m) | 1.00E-08 | 0.006 | 0.396 | 0.575 |
| Mining concession cover (%) | 1.00E-08 | 0 | 0.133 | 0.206 |
| Logging concession cover (%) | 1.00E-08 | 1 | 0 | 0.095 |
| REDD+ jurisdiction or other conservation program(s) cover (%) | 6.64E-05 | 0.052 | 0.791 | 0.297 |
| Travel distance to urban centers (days) | 1.00E-08 | 0.039 | 0.042 | 0.034 |
| Average annual deforestation (ha) | 1.00E-08 | 196.545 | 213.683 | 1378.81 |
| Average cumulative deforestation (ha) | 0.999703 | 932.066 | 942.98 | 5828.416 |
| Annual precipitation (mm) | 1.00E-08 | 1728.75 | 1727.138 | 1516.734 |
| Average annual buffer deforestation (ha) | 1.00E-08 | 2367.125 | 1309.851 | 1538.826 |
| Average cumulative buffer deforestation (ha) | 1.00E-08 | 11344.88 | 6596.921 | 6608.888 |

*Based on five control areas.

Table A21. Covariate balance for Project 1325 from Tanzania.

| Covariate | V-weights | Project area | Synthetic control* | Control set average |
|---|---|---|---|---|
| Average tree cover (%) | 6.20E-09 | 0.51 | 0.288 | 0.256 |
| Protected area cover (%) | 0.619576 | 0.176 | 0.176 | 0.317 |
| Slope (degree) | 6.20E-09 | 3.681 | 1.563 | 1.675 |
| Elevation (m) | 3.10E-07 | 313.465 | 556.702 | 804.17 |
| Mining concession cover (%) | 3.32E-07 | 0.086 | 0.052 | 0.03 |
| Travel distance to urban centers (days) | 6.20E-09 | 0.014 | 0.01 | 0.011 |
| Average annual deforestation (ha) | 1.45E-06 | 314.2 | 319.638 | 431.961 |
| Average cumulative deforestation (ha) | 7.88E-05 | 1689.107 | 1692.053 | 2159.901 |
| Annual precipitation (mm) | 0.380334 | 948 | 948 | 914.264 |
| Average annual buffer deforestation (ha) | 6.20E-06 | 807 | 803.937 | 766.278 |
| Average cumulative buffer deforestation (ha) | 2.26E-06 | 4272.727 | 4186.993 | 3868.108 |

*Based on six control areas.

Table A22. Covariate balance for Project 1897 from Tanzania.

| Covariate | V-weights | Project area | Synthetic control* | Control set average |
|---|---|---|---|---|
| Average tree cover (%) | 0.000103 | 0.5 | 0.325 | 0.258 |
| Slope (degree) | 0.589418 | 5.122 | 5.117 | 2.198 |
| Elevation (m) | 4.72E-05 | 1389.528 | 1319.223 | 934.263 |
| Mining concession cover (%) | 5.89E-09 | 0.044 | 0.001 | 0.016 |
| Travel distance to urban centers (days) | 5.89E-09 | 0.016 | 0.015 | 0.012 |
| Average annual deforestation (ha) | 5.89E-09 | 481.319 | 475.931 | 706.79 |
| Average cumulative deforestation (ha) | 0.377348 | 2574.946 | 2588.453 | 5184.995 |
| Annual precipitation (mm) | 0.011534 | 1101.062 | 1113.8 | 903.495 |
| Average annual buffer deforestation (ha) | 5.89E-09 | 751.375 | 814.131 | 740.75 |
| Average cumulative buffer deforestation (ha) | 0.02155 | 3662.5 | 3713.207 | 5446.848 |

*Based on four control areas.

Table A23. Covariate balance for Project 1900 from Tanzania.

| Covariate | V-weights | Project area | Synthetic control* | Control set average |
|---|---|---|---|---|
| Average tree cover (%) | 0.333299 | 0.1 | 0.1 | 0.156 |
| Protected area cover (%) | 0.333299 | 1 | 1 | 0.581 |
| Slope (degree) | 3.74E-05 | 0.804 | 0.804 | 2.526 |
| Elevation (m) | 3.33E-09 | 1172.476 | 1109.468 | 1068.004 |
| Mining concession cover (%) | 3.33E-09 | 0 | 0.041 | 0.028 |
| Travel distance to urban centers (days) | 0.333299 | 0.014 | 0.014 | 0.014 |
| Average annual deforestation (ha) | 3.33E-09 | 0.441 | 0.778 | 26.448 |
| Average cumulative deforestation (ha) | 6.65E-05 | 4.327 | 5.061 | 241.384 |
| Annual precipitation (mm) | 3.33E-09 | 551.562 | 581.125 | 779.517 |
| Average annual buffer deforestation (ha) | 3.33E-09 | 23.188 | 22.812 | 22.856 |
| Average cumulative buffer deforestation (ha) | 3.33E-09 | 147.75 | 107.062 | 196.002 |

*Based on one control area.

Table A24. Covariate balance for Project 1202 from Zambia.

| Covariate | V-weights | Project area | Synthetic control* | Control set average |
|---|---|---|---|---|
| Average tree cover (%) | 0.002879 | 0.24 | 0.24 | 0.254 |
| Protected area cover (%) | 7.83E-09 | 0.128 | 0.87 | 0.286 |
| Slope (degree) | 3.39E-05 | 4.378 | 2.122 | 1.276 |
| Elevation (m) | 7.83E-09 | 1093.45 | 925.286 | 1179.245 |
| Travel distance to urban centers (days) | 7.83E-09 | 0.015 | 0.015 | 0.013 |
| Average annual deforestation (ha) | 0.050949 | 4.716 | 4.716 | 40.79 |
| Average cumulative deforestation (ha) | 0.162489 | 26.411 | 26.351 | 152.984 |
| Annual precipitation (mm) | 0.783331 | 859.875 | 859.875 | 1070.829 |
| Average annual buffer deforestation (ha) | 0.000318 | 109.875 | 98.997 | 109.381 |
| Average cumulative buffer deforestation (ha) | 7.83E-09 | 530.375 | 328.277 | 398.26 |

*Based on six control areas.

Table A25. Covariate balance for Project 1775-1 from Zambia.

| Covariate | V-weights | Project area | Synthetic control* | Control set average |
|---|---|---|---|---|
| Average tree cover (%) | 1.00E-08 | 0.24 | 0.13 | 0.223 |
| Protected area cover (%) | 1.00E-08 | 0.963 | 1 | 0.451 |
| Slope (degree) | 1.00E-08 | 4.627 | 0.644 | 1.222 |
| Elevation (m) | 1.00E-08 | 659.448 | 1060.101 | 1086.366 |
| Travel distance to urban centers (days) | 1.00E-08 | 0.015 | 0.014 | 0.014 |
| Average annual deforestation (ha) | 1.00E-08 | 55.69 | 82.145 | 544.52 |
| Average cumulative deforestation (ha) | 1 | 459.236 | 587.923 | 3337.871 |
| Annual precipitation (mm) | 1.00E-08 | 904.143 | 832.214 | 991.243 |
| Average annual buffer deforestation (ha) | 1.00E-08 | 282.929 | 300.143 | 283.951 |
| Average cumulative buffer deforestation (ha) | 1.00E-08 | 2214.5 | 2347.929 | 1753.376 |

*Based on one control area.

Table A26. Covariate balance for Project 1775-2 from Zambia.

| Covariate | V-weights | Project area | Synthetic control* | Control set average |
|---|---|---|---|---|
| Average tree cover (%) | 9.75E-09 | 0.26 | 0.592 | 0.247 |
| Protected area cover (%) | 0.017502 | 0.995 | 0.975 | 0.199 |
| Slope (degree) | 9.75E-09 | 2.384 | 1.153 | 1.215 |
| Elevation (m) | 9.75E-09 | 697.335 | 1175.69 | 1202.202 |
| Travel distance to urban centers (days) | 0.974804 | 0.014 | 0.014 | 0.012 |
| Average annual deforestation (ha) | 9.75E-09 | 29.063 | 21.558 | 305.833 |
| Average cumulative deforestation (ha) | 0.007695 | 75.096 | 109.56 | 1808.967 |
| Annual precipitation (mm) | 9.75E-09 | 948.643 | 1214.721 | 1109.714 |
| Average annual buffer deforestation (ha) | 9.75E-09 | 331.786 | 299.896 | 333.088 |
| Average cumulative buffer deforestation (ha) | 9.75E-09 | 2034.071 | 2073.964 | 1936.154 |

*Based on two control areas.

Table A27. Covariate balance for Project 1775-3 from Zambia.

| Covariate | V-weights | Project area | Synthetic control* | Control set average |
|---|---|---|---|---|
| Average tree cover (%) | 9.99E-09 | 0.2 | 0.423 | 0.198 |
| Slope (degree) | 5.76E-07 | 1.186 | 1.043 | 1.095 |
| Elevation (m) | 9.99E-09 | 657.042 | 1182.804 | 1001.683 |
| Travel distance to urban centers (days) | 9.99E-09 | 0.013 | 0.014 | 0.014 |
| Average annual deforestation (ha) | 0.99903 | 11.275 | 11.275 | 106.349 |
| Average cumulative deforestation (ha) | 0.000968 | 59.63 | 60.052 | 685.327 |
| Annual precipitation (mm) | 9.99E-09 | 911 | 1224.422 | 976.04 |
| Average annual buffer deforestation (ha) | 2.04E-08 | 110.714 | 117.908 | 111.107 |
| Average cumulative buffer deforestation (ha) | 1.32E-06 | 678.714 | 672.593 | 706.651 |

*Based on four control areas.

Table A28. Covariate balance for Project 904 from Cambodia.

| Covariate | V-weights | Project area | Synthetic control* | Control set average |
|---|---|---|---|---|
| Average tree cover (%) | 0.000498 | 0.42 | 0.477 | 0.399 |
| Protected area cover (%) | 0.157606 | 0.443 | 0.442 | 0.169 |
| Slope (degree) | 8.35E-09 | 0.869 | 2.147 | 1.477 |
| Elevation (m) | 8.35E-09 | 69.484 | 106.046 | 96.806 |
| Mining concession cover (%) | 0.00411 | 0.153 | 0.116 | 0.125 |
| Soil fertility level | 8.35E-09 | 1.75 | 1.888 | 2.157 |
| Other economic concessions cover (%) | 0.002572 | 0.025 | 0.127 | 0.119 |
| Travel distance to urban centers (days) | 0.000616 | 0.012 | 0.017 | 0.005 |
| Average annual deforestation (ha) | 8.35E-09 | 147.463 | 152.14 | 1163.323 |
| Average cumulative deforestation (ha) | 0.834598 | 467.113 | 482.857 | 4554.692 |
| Annual precipitation (mm) | 8.35E-09 | 1407.286 | 1844.49 | 1824.241 |
| Average annual buffer deforestation (ha) | 8.35E-09 | 2541 | 1755.709 | 2107.064 |
| Average cumulative buffer deforestation (ha) | 8.35E-09 | 8162.571 | 6316.207 | 8230.669 |

*Based on five control areas.

Table A29. Covariate balance for Project 1650 from Cambodia.

| Covariate | V-weights | Project area | Synthetic control* | Control set average |
|---|---|---|---|---|
| Average tree cover (%) | 0.000498 | 0.73 | 0.76 | 0.679 |
| Protected area cover (%) | 0.157606 | 0.97 | 0.732 | 0.677 |
| Slope (degree) | 8.35E-09 | 2.566 | 1.155 | 2.268 |
| Elevation (m) | 8.35E-09 | 197.462 | 86 | 144.59 |
| Mining concession cover (%) | 0.00411 | 0.736 | 0.208 | 0.248 |
| Soil fertility level | 8.35E-09 | 1.714 | 2 | 2.069 |
| Other economic concessions cover (%) | 0.002572 | 0.005 | 0.159 | 0.197 |
| Travel distance to urban centers (days) | 0.000616 | 0.014 | 0.037 | 0.024 |
| Average annual deforestation (ha) | 8.35E-09 | 214.199 | 232.645 | 1338.094 |
| Average cumulative deforestation (ha) | 0.834598 | 993.492 | 1135.273 | 5807.584 |
| Annual precipitation (mm) | 8.35E-09 | 2124.5 | 1800.2 | 1850.923 |
| Average annual buffer deforestation (ha) | 8.35E-09 | 3167.3 | 1343 | 1687.065 |
| Average cumulative buffer deforestation (ha) | 8.35E-09 | 12213.2 | 5575.8 | 7342.892 |

*Based on one control area.